\documentclass{aa}

\usepackage[colorlinks=true,allcolors=blue]{hyperref}
\usepackage[varg]{txfonts}
\usepackage{amsmath}
\usepackage{mathtools}
\usepackage{graphicx}
\usepackage[dvipsnames]{xcolor}
\usepackage{array}
\usepackage{gensymb}  
\usepackage{tablefootnote}
\usepackage{lipsum}

\usepackage{orcidlink}

\begin{document}

\newcommand{\todo}[1]{\textcolor{cyan}{\bf Todo: #1}}
\newcommand{\rev}[1]{{#1}}

\title{
  Warped Disk Evolution in Grid-Based Simulations
}


\author{
	C. N. Kimmig \orcidlink{0000-0001-9071-1508}\inst{1}
	\and C. P. Dullemond \orcidlink{0000-0002-7078-5910}\inst{1}
}

\institute{
	 Zentrum f\"ur Astronomie, Heidelberg University, Albert-Ueberle-Str.~2, 69120 Heidelberg, Germany
}

\date{Received xxx / Accepted xxx} 

\abstract {Multiple observations show evidence that a significant fraction of protoplanetary disks \rev{contain warps}.
A warp in a disk evolves in time affecting the appearance and shape of shadows and arcs.
It also greatly influences kinematic signatures.
Understanding warp evolution helps provide valuable insight to its origin.
}
{So far, many theoretical studies of warped disks have been conducted using Smoothed Particle Hydrodynamics (SPH) methods.
In our approach, we use a grid-based method in spherical coordinates which has notable advantages: the method allows for accurate modelling of low viscosity values and the resolution does not depend on density or mass of the disk, which allows \rev{surface structures to be resolved}.
}
{We perform 3D simulations using FARGO3D to simulate the evolution of a warped disk and compare the results to one-dimensional models using a ring code.
Additionally, we extensively investigate the applicability of grid-based methods to misaligned disks and test \rev{their} dependency on grid resolution as well as disk viscosity.
}
{We find that grid-based hydrodynamic simulations are capable of simulating disks not \rev{aligned to} the grid geometry.
Our three-dimensional simulation of a warped disk compares well to one-dimensional models in evolution of inclination.
However, we find a twist which is not captured in 1D models.
After thorough analysis we suspect this to be a physical effect possibly caused by non-linear effects neglected in the one-dimensional equations.
Evaluating the internal dynamics, we find sloshing and breathing motions as predicted in local shearing box analysis.
They can become supersonic, which may have strong consequences on kinematic observations of warped disks.
}
{Warped disks can be accurately modelled in three-dimensional grid-based hydrodynamics simulations when using reasonably good resolution, especially in \rev{the} $\theta$-direction.
We find good agreement with the linear approximation of the sloshing motion which highlights the reliability of 1D models.
}

\maketitle

\section{Introduction}

As more and more observations of protoplanetary disks become available \citep[][etc.]{Andrews2018, Andrews2020, Garufi2022}, it is astonishing how many of them show non-axisymmetric features such as bright arcs as in HD 143006 \citep{Perez2018} or HD 135344B \citep{Cazzoletti2018}, spirals as in MWC 758 \citep{Benisty2015}, narrow shadow lanes as in HD 142527 \citep{Marino2015} or in HD 135344B \citep{Stolker2017}, or wide shaded regions as in TW Hydra \citep{Debes2017} or in HD 139614 \citep{Muro-Arena2020}.
Some of these asymmetrical features, in especially shadow lanes and wide shaded regions, can be traced back to a misaligned part of the disk \rev{casting a shadow.} 

Several misaligned disks have been detected. \rev{A few further examples of such misaligned disks are} HD 142527 \citep{Casassus2015}, HD 100453 \citep{Benisty2017}, HD 143006 \citep{Benisty2018}, DoAr 44 \citep{Casassus2018}, GW Orionis \citep{Kraus2020, Smallwood2021}, KH 15D \citep{LodatoFacchini2013, Poon2021}, and many more.
\cite{Bohn2022} specifically targeted disks with shadow features and measured the misalignment angle between inner and outer \rev{regions} of the disk by looking at each disk in near infrared observations to probe the inner disk and \rev{submillimeter} interferometric observations for the outer disk.
Out of a sample of 20 disks, they find six disks with a significant misalignment and five disks showing no misalignment, while the remaining nine disks are not analysable with certainty using the current data.
\rev{A study by} \cite{Min2017} involves analytical considerations of shadow locations to deduce inner disk geometries.
To investigate the shadow morphology, \cite{Nealon2019} studied shadows in synthetic scattered light observations.
\cite{Nealon2020shadows} found that shadows cast by a misaligned inner disk can move in time and appear to rock back and forth depending on the misalignment angle and other disk geometry properties.

\rev{Misaligned} disks can be classified into so-called \textit{broken disks} that have two or more parts of the disk misaligned with respect to each other, and \textit{warped disks}, that contain a continuous misalignment which means that all parts of the disk is connected and the inclination angle changes with radius.
The classification is, however, not perfectly sharp as there can be objects falling in between those categories or be a mixture of both.

There are multiple mechanisms leading to the formation of misaligned disks.
In all cases, a misaligned component, either external or internal or both, is exerting a torque on the disk.
In one scenario, a binary misaligned with respect to the disk plane, gravitationally torques the disk.
The misalignment can form in a circumprimary or -secondary disk with an outer companion \citep{Papaloizouetal1995, Zanazzi2018a}, as well as in a circumbinary disk \citep{Facchini2013, LodatoFacchini2013, Zanazzi2018b, DengOgilvie2022}.
In the case of an inclined outer companion, the disk can under certain conditions precesses as a rigid body \citep{PapaloizouTerquem1995, Papaloizou1997} or break \citep{Dogan2023, Rabago2023b}.
An inner companion can drive a warp leading to alignment of the disk with the binary orbital plane \rev{\citep{FoucartLai2013}} or polar alignment \citep{Zanazzi2018b, Rabago2023}.
The disk can also get warped if the companion is not bound to the system but passing by in a so-called fly-by event \citep{ClarkePringle1993, Cuello2018, Nealon2020flyby}.
A misaligned planet can also cause a warp \citep{Nealon2018}.
This scenario is similar to the misaligned binary case, however with much smaller masses of the perturber.
Such a misaligned planet could have been captured by the star-disk system or kicked out of the disk plane by N-body interaction with other objects such as other planets or fly-by objects \citep{Nagasawa2008}.
Late infall of material onto a disk could be another cause for a warp \citep{Kueffmeier2021}.
In this scenario, the infalling material adds angular momentum with a different direction to the disk.
Warped disks in combination with infalling material have also been observed \citep{Sai2020}.
Warping of disks can affect planet formation \citep{Aly2021}.

In warped disks, pressure forces occur that act to change the disk's orbital plane which means that a warped disk without a component driving the warp is not static but evolves in time \rev{\citep{PapaloizouLin1995, PapaloizouPringle1983}}.
The evolution occurs in different manners depending on the disk properties, i.e. its viscosity described by the Shakura-Sunyaev turbulence parameter $\alpha_\mathrm{t}$ \citep{Shakura1973} and its thickness, or more specifically its aspect ratio $h = h_\mathrm{p}/r$, where $h_\mathrm{p}$ is the pressure scale height of the disk and $r$ the distance from the star.
Disks with high viscosity compared to the aspect ratio $\alpha_\mathrm{t} > h$ evolve in a diffusive manner.
In this regime, the warp is directly dampened and dissipates \citep{Pringle1992, Ogilvie1999, LodatoPrice2010}.
Disks with lower viscosity, i.e. $\alpha_\mathrm{t} < h$, fall in the so-called wave-like regime.
In this regime, the warp travels as a wave through the disk with a wave speed of $c_\mathrm{s}/2$ \citep{LubowOgilvie2000, Gammie2000, OgilvieLatter2013a, OgilvieLatter2013b}.
\rev{Over time, the warp is dampened as well in this regime, if $\alpha_{t} \neq 0$.}
Because protoplanetary disks are known to have low viscosities of $\alpha_\mathrm{t} = 10^{-5}\text{--}10^{-3}$ \citep{Manara2022PPVII}, they typically fall into the wave-like regime.

Historically, each evolution regime was described by a different set of equations for the warp evolution.
\cite{Martin2019} were able to combine both equation sets to a generalized formalism.
In their generalized set of equations, they included a damping term in order to suppress unphysical features.
This damping term causes the equations to become stiff and challenging to solve numerically.
\citet{Dullemond2022} (hereinafter referred to as DKZ22) were able to derive the same set of equations from shearing box considerations and found the reason for these unphysical features, as the equations need to take the rotation of the orbital plane into account.
They proposed an alternative, physically motivated term to correct for the orbital plane rotation.
The generalized equation set includes mass and angular momentum conservation, both of which depend on the internal torque vector \vec{G}.
\begin{equation} \label{eq:mass-conservation}
\frac{\partial \Sigma}{\partial t} - \frac{2}{r} \frac{\partial}{\partial r}\left[ \frac{\frac{\partial(r \vec{G})}{\partial r} \cdot \vec{l}}{r \Omega} \right] = 0
\end{equation}
\begin{equation} \label{eq:angular-momentum-conservation}
\frac{\partial \vec{L}}{\partial t} - \frac{2}{r} \frac{\partial}{\partial r}\left[ \frac{\frac{\partial (r\vec{G})}{\partial r} \cdot \vec{l}}{\Sigma r \Omega} \ \vec{L} \right] + \frac{1}{r} \frac{\partial (r\vec{G})}{\partial r} = T
\end{equation}
Here, $\Sigma$ is the surface density, $\Omega$ the Kepler orbital frequency and $\vec{L} = \Sigma R^2 \Omega \vec{l}$ the angular momentum density with $\vec{l}$ as angular momentum unit vector.
The internal torque vector $\vec{G}$ arises due to the misalignment of neighbouring orbits (DKZ22) and takes the form
\rev{
\begin{equation} \label{eq:internal-torque}
\begin{split}
\frac{\partial \vec{G}}{\partial t} + \left(\frac{\kappa^2 - 1}{2}\right) \Omega \ \vec{l} \times \vec{G} & + \alpha_\mathrm{t} \Omega \vec{G} + \left( \vec{l} \times \frac{\partial \vec{l}}{\partial t} \right) \times \vec{G} \\
 = & - \frac{\Omega^3 r \Sigma h_\mathrm{p}^2}{4} \ \frac{\partial \vec{l}}{\partial \ln r} + \frac{3}{2} \alpha_\mathrm{t}^2 \Omega^3 r \Sigma h_\mathrm{p}^2 \vec{l},
\end{split}
\end{equation}
}
where \rev{ $\kappa = \Omega_\mathrm{e} / \Omega$ with $\Omega_\mathrm{e}$ as epicyclic frequency}.
Note that we are using the definition of the internal torque vector $\vec{G}$ and the notation following DKZ22, which differs from the definition of other works, e.g. \cite{Martin2019} or \cite{Nixon2016} by a \rev{factor of $-1/r$} (see DKZ22, Appendix C).

As mentioned, the internal torque arises due to the misalignment of neighbouring orbits.
More specifically, annuli are deformed (see DKZ22, Figure~9) and therefore act a pressure on neighbouring annuli.
This leads to a dynamic surface where material flows radially inward and outward within one orbit.
This motion is called sloshing motion (DKZ22), often referred to as ``resonant motion'' in literature \citep{Nixon2016}.
The sloshing motion is the main drive for warp evolution.

Warped disks can be investigated in one-dimensional as well as three-dimensional models.
In the one-dimensional models, the disk is to split into concentric annuli.
Each annulus plane can be described by its normal vector, which is identical to the angular momentum unit vector.
This angular momentum unit vector can then be updated following equation~\ref{eq:angular-momentum-conservation}.
While one-dimensional models have a huge advantage in regard to computation time, they \rev{are obviously simplified.}
Additionally, the equations are derived in a linear approximation and contain additional assumptions, e.g. about their Keplerity.
Therefore, it is useful to model warped disks in full three-dimensional hydrodynamic simulations.

Three-dimensional simulations of warped disks are often performed using smoothed particle hydrodynamics (SPH) models \citep{GingoldMonaghan1977}.
However, the SPH method is usually implemented with an artificial viscosity that allows for shock treatment \citep{Monaghan2005}.
Simulations of disks modelled in SPH need to have a disk viscosity higher than the artificial viscosity in order to capture the physical processes accurately.
Protoplanetary disks, however, are known to have a rather low viscosity \citep{Pinte2016, Villenave2020}, which means that a method able to model lower viscosities would be more fitting.
Additionally in the most common SPH models, the resolution highly depends on density.
For a protoplanetary disk this means that the surface of the disk is always less resolved than its midplane.
For warped disks, however, it is advantageous to resolve the surface structures and dynamics well due to the sloshing motion mechanism.
More importantly, observations of protoplanetary disks often trace the surface layers, which is why accurate model predictions for these regions are important.

In this work, we model warped disks in full three-dimensional hydrodynamics using a grid-based code.
This way, we are able to model protoplanetary disks with low viscosities, without the resolution dependency on density or any other physical quantity.
However, we are aware that a grid can also cause numerical effects, as for example found by \cite{Hopkins2015} (Figure 8 therein).
In order to minimize them, we choose a spherical symmetry of the grid to take advantage of the disk's symmetry.
As we still expect some numerical effects caused by the grid, we carefully test the \rev{set-up} in the test case of a planar and smooth disk inclined with respect to the grid geometry.
Physically, this should give the same result as a planar disk \rev{aligned to} the grid geometry and allows us to compare and extract the grid effects.
A few studies used grid-based methods to model disks with out-of-plane features before \citep{FragnerNelson2010, Rabago2023b, Rabago2023}.

This paper is organized as follows. We describe our numerical \rev{set-up} in Section~\ref{sec:numerical-methods} and the investigation of the grid effects in the case of a disk tilted with respect to the midplane in Section~\ref{sec:grid-effects}. We present and discuss the results of a warped disk simulation in Section~\ref{sec:warpeddisk}. In Section~\ref{sec:sloshing}, we investigate the internal torque and sloshing motion in the 3D simulation. We conclude this work in Section~\ref{sec:conclusion}.

\section{Numerical Method} \label{sec:numerical-methods}

We perform three-dimensional hydrodynamic simulations using the versatile code FARGO3D by \cite{FARGO3D}.
In our simulations, we use the implemented hydrodynamics modelling a system consisting of a central star and a disk surrounding it.
For our purposes, we do not consider magnetic fields and therefore do not use the magnetohydrodynamic features of FARGO3D.
We further do not include any planets or other companions.

\begin{figure} [ht!]
  \centerline{\includegraphics[width=0.25\textwidth]{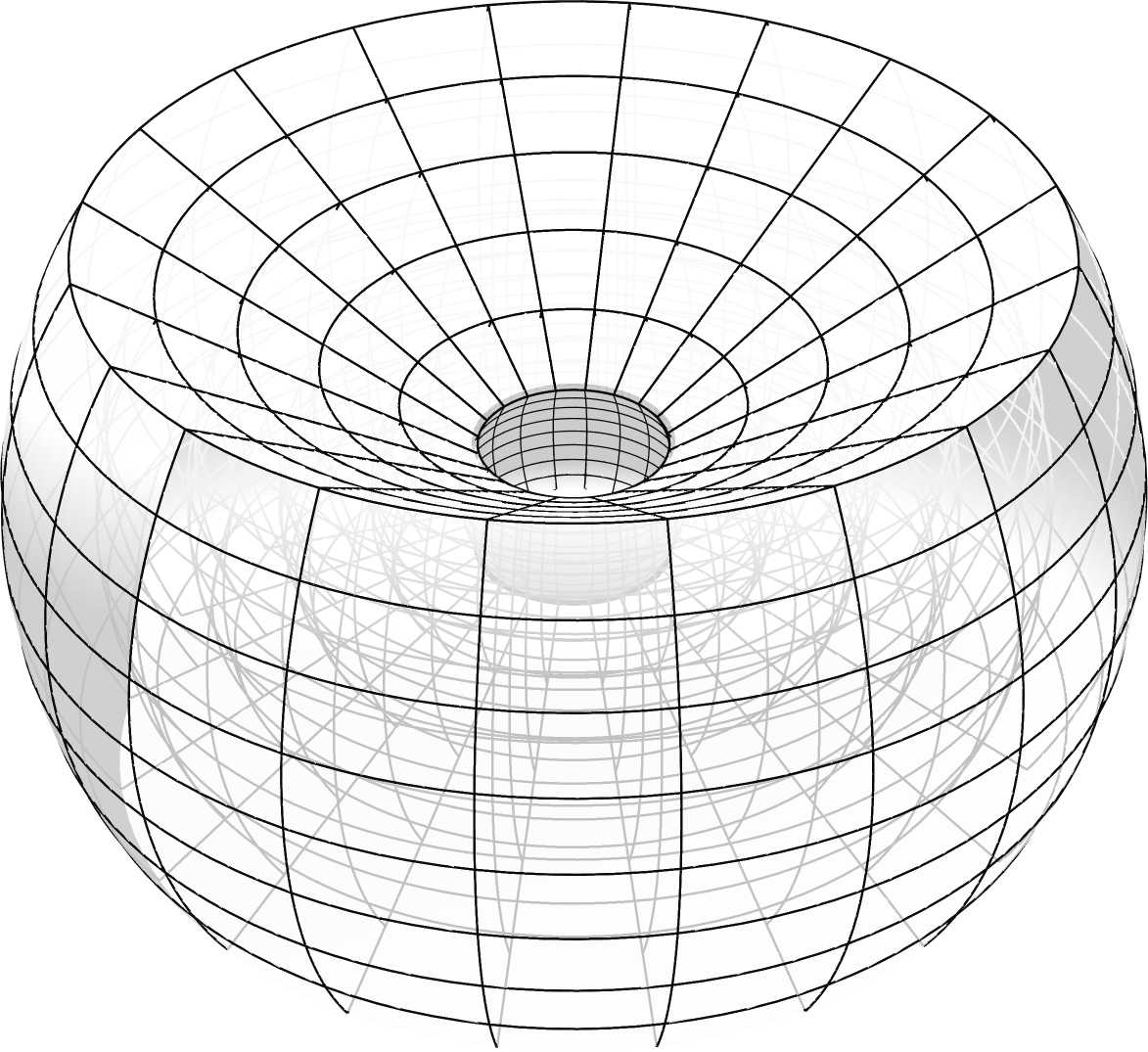}}
  \caption{\label{fig:spherical-coord} Schematics of the spherical grid \rev{set-up} in our simulations. The poles are cut off in order to save computation time.
  }
\end{figure}

In order to take advantage of the system's symmetry, we set up a model in spherical coordinates. To save computation time, we restrict the vertical computation domain to a certain range, meaning that the poles of the grid sphere are cut off (see Figure~\ref{fig:spherical-coord}).
Because we are planning to look at features tilted with respect to the midplane of the grid, we need to consider a large enough regime in the vertical direction.

We set up a protoplanetary such that the surface density follows the power law
\begin{equation}
\Sigma(r) = \Sigma_0 \left( \frac{r}{R_0} \right)^{-p}
\end{equation}
with slope $p$ and the reference scale $R_0$. In FARGO3D, this reference scale is usually set to $R_0~=~5.2\,\mathrm{au}$, which we adopt in our models.
\rev{In general, we} assume a locally isothermal disk, which means that the temperature structure does not change in time.
However, the temperature may very well vary within the disk and is not constant as a function of distance to the star.
We set the temperature structure so that the disk is flared, such that the aspect ratio can be described as
\begin{equation} \label{eq:aspectratio}
h(r) = h_0 \left( \frac{r}{R_0} \right)^{i_\mathrm{fl}},
\end{equation}
where $h_0$ is the unflared aspect ratio and $i_\mathrm{fl}$ the flaring index. \rev{For most of our simulations, we adopt a globally isothermal model by choosing $i_\mathrm{fl}=0.5$.}
The initial velocity in azimuthal direction is set to the Keplerian velocity, with a correction term for pressure gradients.

In our simulations, we use reflective boundary conditions in the radial and vertical direction, i.e. we do not allow inflow and outflow.
This gives the most control over the physics at the boundary, however, is not very natural, therefore this condition should be relaxed in further work.
We use the implemented wave-killing boundary conditions \citep{Val-Borro2006} in order to damp wave reflections, where the damping zones have a width of 10\% of the grid edge radius.
We intentionally do not choose boundary conditions fixing the ghost cells to an analytical extrapolation, as this is not possible for the warped disk case.

Aiming to investigate the capability of the code to simulate features out of the grid plane, we set up a planar disk tilted with respect to the midplane of the grid.
For the rotation of the disk, we use the analytical initial disk model of the FARGO3D code and incline it using appropriate coordinate transformations.
This way, we achieve a disk tilted with respect to the grid midplane (see Section \ref{sec:tilteddisk}).

For the warped disk model, we use the same coordinate transformation, only now using a tilt angle depending on distance from the star instead of a constant tilt angle. The function we use is inspired by \citet{Martin2019} (Eq. 19 therein) and reads
\begin{equation}\label{eq:incl-ini}
i(r) = i_\mathrm{max} \left[\frac{1}{2} \tanh \left( \frac{r - r_\mathrm{warp}}{r_\mathrm{width}} \right) + \frac{1}{2} \right],
\end{equation}
where $i_\mathrm{max}$ is the maximum warp tilt, $r_\mathrm{warp}$ the location of the steepest warp, and $r_\mathrm{width}$ the width of the warp transition.
Depending on the disk tilt or warp maximum inclination, we need to adjust the grid space in the vertical direction.

\section{Investigating the Grid Effects} \label{sec:grid-effects}

In this part, we aim to investigate the feasibility of simulating a warped disk using a grid-based method, in our case FARGO3D \citep{FARGO3D}.
A warped disk includes features that do not align with the grid geometry, specifically they are tilted with respect to the grid midplane.
Such features can increase numerical friction compared to features perfectly aligning with the grid geometry.
In order to investigate the numerical influences of the grid on the out-of-plane features, we first investigate the case of a planar disk tilted with respect to the grid midplane.
Physically, this is the same scenario as a planar disk \rev{aligned to} the grid midplane.
Comparing a tilted disk to an untilted disk allows us to extract the differences between both cases, which can only be caused by the misalignment between disk and grid midplanes.
This way, we achieve a gauge for the grid effects.

In the untilted scenario, we expect standard viscous evolution.
For low viscosities, the time scale for viscous evolution is long enough so that we do not expect the surface density to change much when looking at the short-term evolution.
In an ideal \rev{set-up}, a disk tilted with respect to the grid midplane should evolve in the same way.
Intuitively, one might think the spherical coordinate system is spherically symmetric and therefore would not numerically influence the \rev{set-up}.
However, a spherical grid always includes directions, i.e. an equator and poles.
This means that \rev{because the orbital motion in a tilted disk is not
aligned with the coordinate system}, a gas parcel's orbit vertically travels through several grid layers: the orbit is below the grid midplane one side and above on the other.
This results in an additional numerical friction influencing the simulation result.
\rev{A similar study is qualitatively mentioned by \cite{FragnerNelson2010}, who find that the numerical viscosity can drag the disk toward alignment.}

\subsection{Untilted Reference Case} \label{sec:reference-case}

\begin{figure} [ht!]
  \centerline{\includegraphics[width=0.5\textwidth]{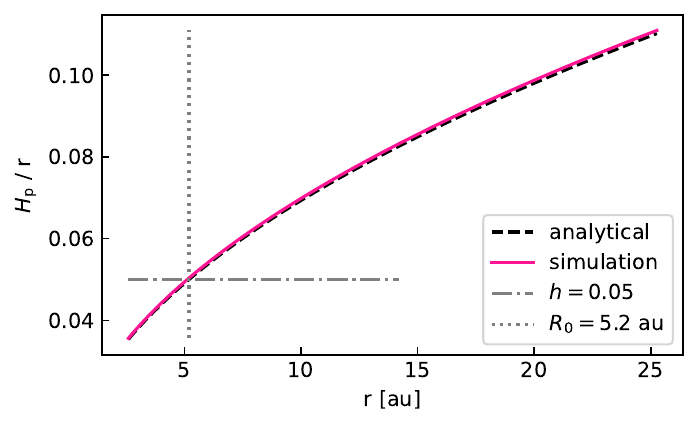}}
  \caption{\label{fig:aspectratio} \rev{Aspect ratio (i.e., pressure scaleheight scaled with radius) $h(r) = H_\mathrm{p}(r) / r$ of our initial set-up. The black dashed line is the analytical calculation according to Equation~\ref{eq:aspectratio}, the pink solid line shows the Gaussian fit from the initial set-up in our simulation. Gray dotted and dashed lines indicate the reference radius $R_0=5.2\,\mathrm{au}$ and the corresponding reference aspect ratio $h_0 = 0.05$, respectively. }}
\end{figure}

We first model the reference case of an untilted disk.
We set up a disk with mass $M_\mathrm{disk} = 0.02\,\mathrm{M_\odot}$ around a Solar-like star with a viscosity of $\alpha_\mathrm{t} = 10^{-3}$. The surface density parameter is set to $\Sigma_0~=~26.2\,\mathrm{g/cm}^2$ with a power law exponent of $p = 1$.
Note that in FARGO3D, the \rev{set-up} is scalable so that the results also apply to different mass and length scales.
The disk ranges from $r_\mathrm{in}=2.6\,\mathrm{au}$ to $r_\mathrm{out}=26\,\mathrm{au}$.
We choose \rev{an aspect ratio at $R_0$ of $h_0=0.05$ and} a flaring index of $i_\mathrm{fl}=0.5$ and set the temperature structure accordingly.
\rev{In Figure~\ref{fig:aspectratio}, we present the aspect ratio of the initial set-up, with both the analytical calculation according to Equation~\ref{eq:aspectratio} (black dashed line) and a fit from our simulation data (pink solid line).
For the fit, we used a Gaussian function to fit the vertical density profile at each radius according to
\begin{equation}
\rho (z) = \rho_0 \exp \left(-\frac{z^2}{2 H_\mathrm{p}^2} \right),
\end{equation}
where $\rho_0$ is the midplane density, $z$ the height above the midplane, and $H_\mathrm{p}$ the pressure scaleheight of the disk.
We note that for simplicity, we used the radial shells of the grid instead of extrapolated cylindrical radii to fit the Gaussian profiles. This means that the fitted (pink) curve is only an approximation, however, sufficient for the purpose of double-checking our initial set-up.
The parameters we chose, i.e., the flaring index of $i_\mathrm{fl} = 0.5$, results in a globally isothermal disk.
}

We run the tilted disk simulations for 500 orbits at $R_0~=~5.2\,\mathrm{au}$, which translates to $t_\mathrm{tot}=6,000\,\mathrm{yr}$ with one orbit being $t_\mathrm{orbit}~=~12\,\mathrm{yr}$\footnote{Note that one actual orbit at $5.2\,\mathrm{au}$ is $11.858\,\mathrm{yr}$, which we round to $12\,\mathrm{yr}$ for simplicity.}.
For now, we set the resolution of 80 cells in $r$-direction with the grid logarithmically spaced in $r$, 100 cells in azimuthal $\varphi$-direction and 40 cells in vertical $\theta$-direction with a range of $-25.8\degree<\theta<25.8\degree$.

\begin{figure} [ht!]
  \centerline{\includegraphics[width=0.5\textwidth]{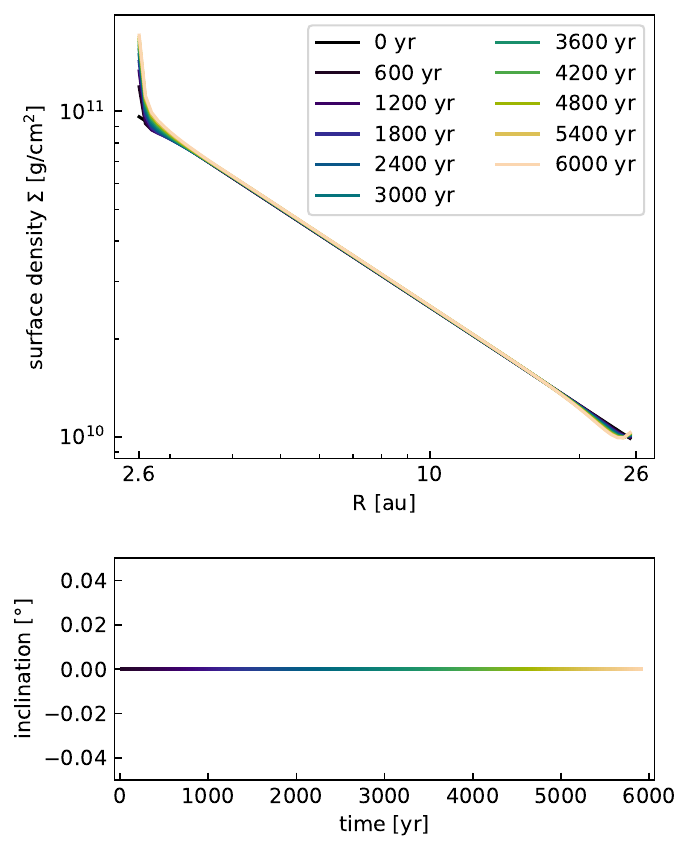}}
  \caption{\label{fig:0deg} Surface density (top panel) and inclination (bottom panel) evolution of a disk \rev{aligned to} the grid midplane. The color in the bottom panel highlights the time. The simulation covers a time span of 6000 years, which are 500 orbits at the reference radius $R_0 = 5.2\,\mathrm{au}$.}
\end{figure}

From the three-dimensional simulations, we compute the surface density using
\begin{equation}
\Sigma(r)  = \int^{\pi}_0 \rho_\varphi (r, \theta)\ r\ \sin(\theta)\ \mathrm{d}\theta,
\end{equation}
where $\rho_\varphi(r, \theta)$ is the azimuthally averaged density of the disk.
Figure \ref{fig:0deg} shows the time evolution of the surface density (top panel) and the inclination (bottom panel), defined as the angle between the angular momentum vector and the z-axis.
A detailed description of how we retrieve the inclination can be found in Section~\ref{sec:tilteddisk}.

As expected, the inclination remains zero throughout the whole simulation which means the disk stays aligned to the grid midplane.
The surface density stays constant on our time scales, except for some deviations at the grid inner and outer edge, due to the rigid boundary conditions.
Because this is merely a reference case, we do not investigate the deviations in detail.

\subsection{Tilted Disk}\label{sec:tilteddisk}

\begin{figure} [ht!]
  \centerline{\includegraphics[width=0.5\textwidth]{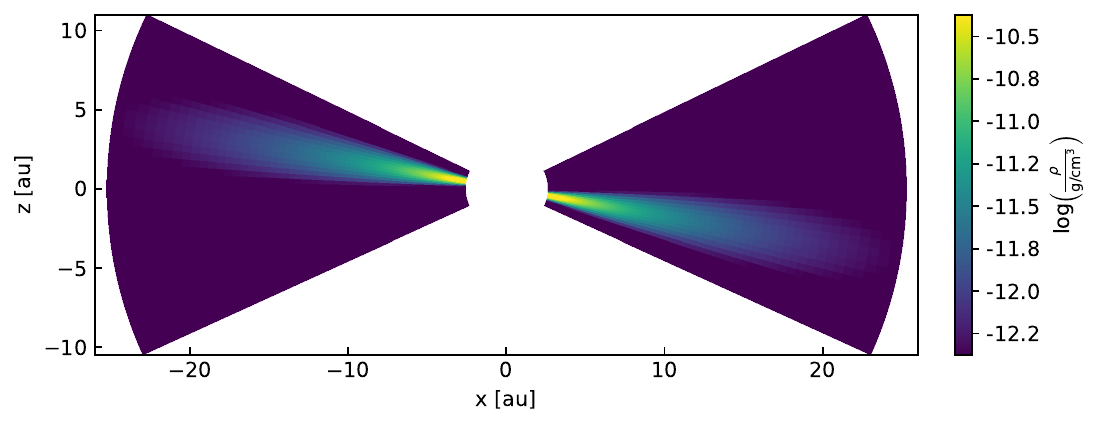}}
  \caption{\label{fig:10deg-crosssec} \rev{Cross-section} through the initial tilted disk along the $x$-axis.}
\end{figure}

We now tilt the disk by ten degrees with respect to the grid midplane.
In a first test, we use the same resolution as in the reference case \rev{aligned to} the midplane.
We set up the tilt in $x$-direction, meaning that the total angular momentum vector of the disk points in direction of the $x$-axis, which means a rotation of the disk around the $y$-axis.
Figure \ref{fig:10deg-crosssec} shows a \rev{cross-section} of this \rev{set-up}.

In order to retrieve the inclination angle, we determine the total angular momentum vector for each radial shell $\vec{L}_r$.
To achieve this, we transform the cell coordinates, as well as the gas velocities, from spherical to Cartesian coordinates.
We can then calculate the angular momentum vector of each shell by
\rev{
\begin{equation} \label{eq:angmom-simulation}
\vec{L}_r = \int \rho(r, \theta, \phi) \ \vec{r} \times \vec{v} \ r^2 \sin(\theta) \mathrm{d}\theta \mathrm{d}\phi,
\end{equation}
}
\rev{where $\vec{v}$ the gas velocity}.
Note that the resulting $\vec{L}_r$ is given in Cartesian coordinates as well.
By normalizing $\vec{L}_r$, we get the direction of angular momentum $\vec{l}_r$, which we use to compute the inclination angle $i$ using
\begin{equation}
\rev{i = \arccos \left(\vec{l} \cdot \vec{l}_z \right)},
\end{equation}
where $\vec{l}_z$ is the unit vector of the $z$-axis with $(0,0,1)$.

\begin{figure} [ht!]
  \centerline{\includegraphics[width=0.5\textwidth]{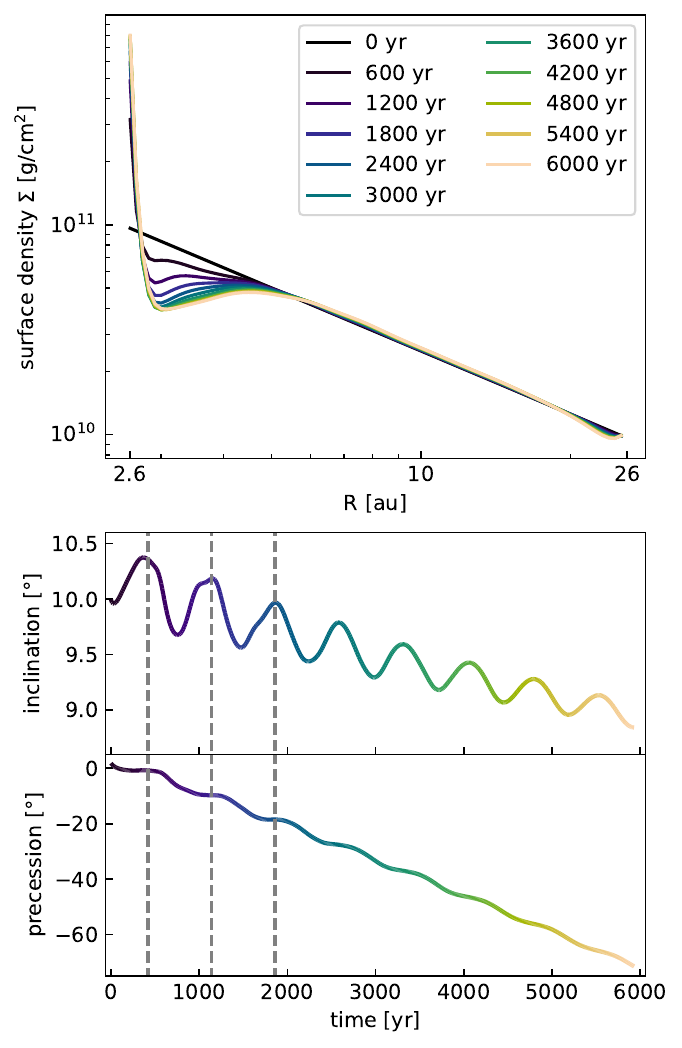}}
  \caption{\label{fig:10deg} Similar to Figure \ref{fig:0deg}, but for the a disk tilted by 10$\degree$ with respect to the grid midplane. The additional panel at the bottom shows the precession angle of the disk. The grey dashed lines indicate the inclination oscillation periods investigated in Figure \ref{fig:wobbles}.}
\end{figure}

In addition to the inclination, we investigate the precession angle, which we define as the clockwise angle between angular momentum vector projected onto the $x$-$y$-plane and the $x$-axis.
This definition gives an initial precession angle of $0\degree$.

\begin{figure} [ht!]
  \centerline{\includegraphics[width=0.5\textwidth]{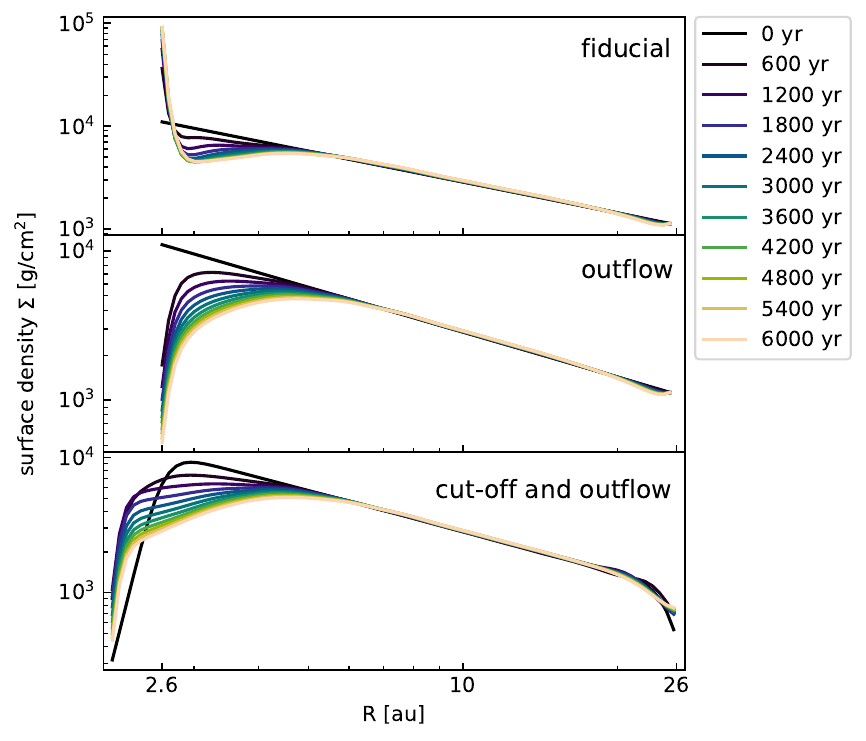}}
  \caption{\label{fig:tilt-surfdens} \rev{Surface density evolution of a tilted disk by $10\degree$ in simulations with different boundary conditions. The top panel shows our fiducial set-up with rigid boundaries, the middle is the same set-up, but outflow boundary conditions at the inner edge, and the bottom panel has an exponential cutoff at outer and inner edge in addition to inner outflow boundary conditions. } }
\end{figure}

Figure \ref{fig:10deg} shows the surface density evolution, as well as the evolution of inclination and precession angle.
The surface density shows a stronger pile-up at the inner boundary and an adjacent density deficit, which we attribute to effects caused by the reflective (rigid) boundary.
\rev{We investigate this by running two more simulations of the same tilted disk set-up, but now using outflow boundary conditions at the inner edge in radial direction, shown in Figure~\ref{fig:tilt-surfdens}. Apart from the boundary condtitons, one simulation has the exact same initial conditions, while the other is set up with an exponential cut-off at the inner and outer disk edges.
In this comparison, we indeed see that the pile-up at the inner edge is caused by the rigid boundary conditions and does not occur in the simulations with inner outflow conditions. However, we observe that the inner region of the disk shows a density deficit in comparison to the initial state ranging up to $5\,\mathrm{au}$. Interestingly, this seems to occur in all three simulations, which suggests that this is not an effect caused by the inner boundary.
We suspect that this effect could be caused by an unphysical warp that we find in the simulation, which we discuss later in this section. }
As we are mainly interested in the alignment with the grid, we do not \rev{go into more detail here.
We merely note that special care should be taken when looking at the surface density evolution of the inner region  in our simulations. }

In an ideal \rev{set-up}, the inclination and precession angle should stay constant.
In our simulation \rev{presented in Figure~\ref{fig:10deg}}, we see that the inclination decreases over time.
This means that the disk is dragged by numerical effects towards alignment with the grid geometry.
However, considering the total decrease of less than 1.5 degrees over a time span of 6000 years, the result is surprisingly good.

The precession angle, however, shows a much stronger trend.
During the simulated 6000 years, the disk precesses by approximately $75\degree$ in a counter-clockwise motion (negative precession angle).
Physically, the disk should not precess, so this is an effect caused by the grid.
The reason for the effect could be numerical friction a gas particle experiences when travelling from one cell to another.
Because the disk is tilted, a gas parcel has to not only travel through all cells in $\varphi$-direction within one orbit, but additionally through several layers in $\theta$-direction.
This increases the amount of numerical friction and might cause the gas parcel to be slightly delayed in its motion through the $\theta$-layers.
Therefore, \rev{a gas parcel arrives later at its} original $\theta$-height and the whole disk shows a precession motion \rev{in the same direction as the orbital rotation, i.e., in our simulation, the orbital direction of the disk is counter-clockwise, as well as the precession.
This can indeed be explained by a delay in $\theta$:
The vertical motion of the gas parcels in a tilted disk during one orbit can be considered like shifting a Gaussian function up and down, similar to an advection.
For advection schemes, \citet{Icke1991} investigated the phase-shift $P$ due to numerical friction (see their Equation 3.10). For $P>0$, the phase is lagging, which means that the transported function is trailing behind the true solution. \citet{Icke1991} found that the investigated higher-order advection schemes resulted in this phase-lag, i.e., $P>0$. Even though FARGO3D uses a different advection algorithm, it is likely this algorithm has the same problem for too low resolution.
Our findings agree with this hypothesis, i.e., we are able to explain the precession we find in our simulation with a phase-lag caused by numerical friction.}
Because the numerical friction decreases when the resolution is increased, we expect the precession effect to be reduced in higher resolutions.
We test this in Section \ref{sec:tilteddisk-resolution}.

In Figure \ref{fig:10deg}, in addition to the decrease of inclination we observe an oscillation pattern.
To investigate the cause of this in detail, we take a look at the inclination profiles in different time steps.
We choose the first and second period from maximum to maximum, which is indicated by the grey dashed lines in Figure \ref{fig:10deg}, bottom panel.
The first period roughly extend from 420 to 1140 years, the second from 1140 to 1860 years.

\begin{figure} [ht!]
  \centerline{\includegraphics[width=0.5\textwidth]{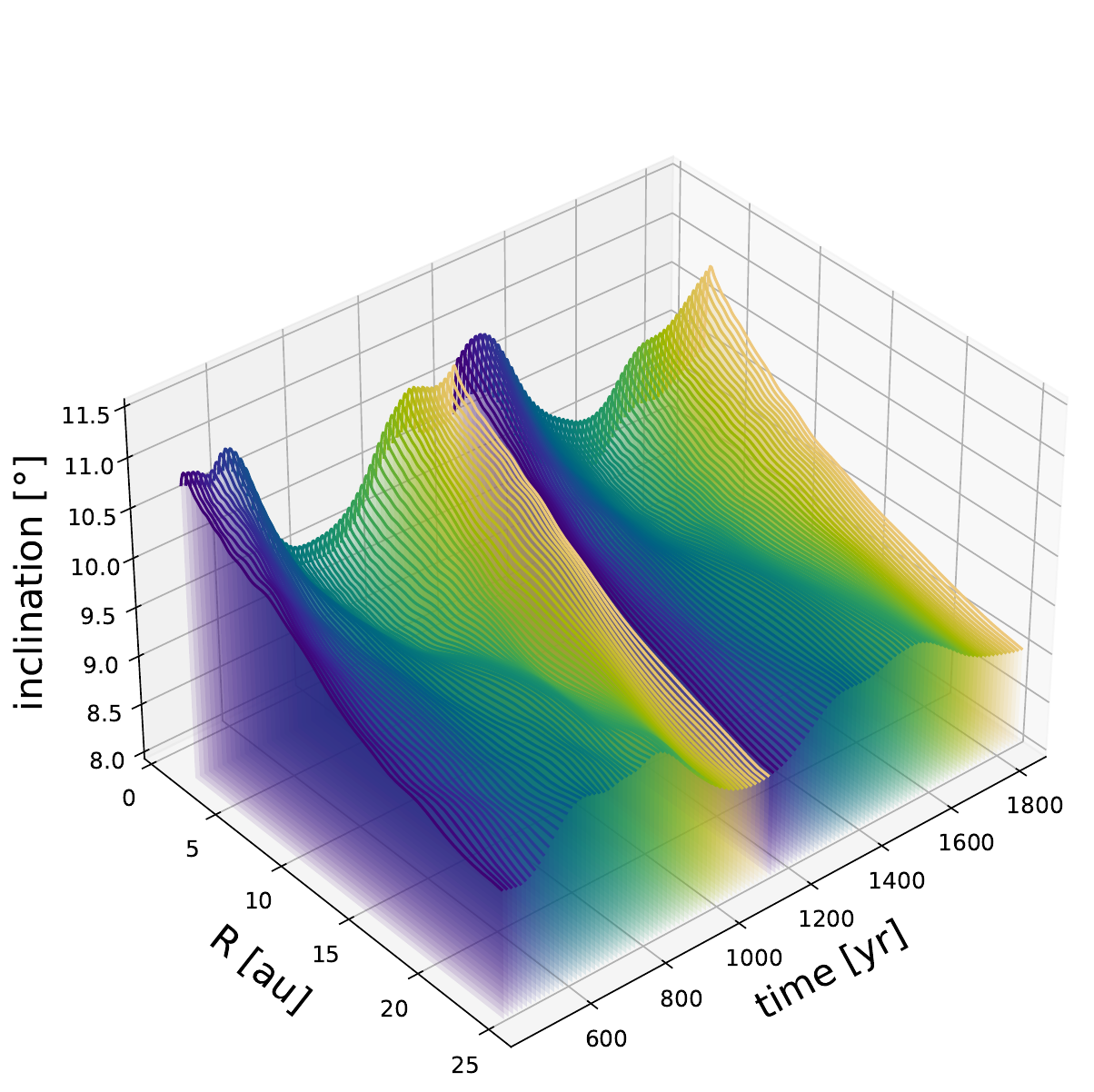}}
  \caption{\label{fig:wobbles} Inclination profile evolution during the first and second periods of the inclination oscillation. Each period is indicated by the color scale from blue to yellow. The first period extends from 420 to 1140 years, the second from 1140 to 1860 years.
  }
\end{figure}

Figure \ref{fig:wobbles} shows the evolution of the inclination profile during these two periods with the time axis as a third dimension, where the color scheme from blue to yellow indicates one period.
We find that the inclination profile varies periodically, which can explain the oscillations in the averaged inclination angle for the total disk.
We find that the disk is not perfectly planar which would have a constant inclination profile with radius.
This means that the disk in our simulation is slightly warped and evolves in a typical wave-like warp evolution.
We note that the warping is modest and not caused by a physical effect but results from numerical grid effects.
The warp might arise due to the grid effects not having the same strength at all radii.
Because the orbital period is shorter at smaller radii, the inner disk experiences more numerical friction than the outer disk in a given time.
As this is not a physical effect, we do not further investigate the evolution of the warp.
Additionally, it is rather modest and does not strongly interfere with our evaluation of the grid effects.
However, it can explain the oscillations in the mean inclination of the total disk.
\rev{We additionally suspect this to influence the evolution of the surface density, as mentioned earlier in this section.}

\subsection{Different Resolutions} \label{sec:tilteddisk-resolution}

In this section, we investigate the grid effects in different resolutions.
We aim to probe the influence of resolution in the three directions and therefore set up three simulations: one with resolution doubled only in $r$-direction (we call this simulation \texttt{r\_resolution}), a second one with resolution doubled in $\varphi$-direction (called \texttt{phi\_resolution}), and a third one doubling only the $\theta$ resolution (\texttt{theta\_resolution}).
In a fourth simulation (\texttt{high\_resolution}), we use a higher resolution in all three directions: 160 cells in $r$, 200 cells in $\varphi$, and 160 cells in $\theta$.
An overview over the simulation resolutions is presented in Table~\ref{tab:resolutions}, with the differences between the simulations highlighted in orange.

\setlength\extrarowheight{5pt}
\begin{table}[h]
\caption{Resolutions in Different Simulations} \label{tab:resolutions}
\begin{tabular}{ | m{3cm} | m{1.4cm}| m{1.4cm} | m{1.4cm} |}
  \hline
  simulation name & cells in $r$ & cells in $\varphi$ & cells in $\theta$ \\
  \hline
  \hline
  fiducial (from Section~\ref{sec:tilteddisk}) & 80 & 100 & 40 \\
  \rev{\small{cells-per-scaleheight (cps) at $r=5.2\,\mathrm{au}$}} & \rev{\small{1.7 cps}} & \rev{\small{0.7 cps}} & \rev{\small{2.2 cps}} \\
  \hline
  \texttt{r\_resolution} & 160 & 100 & 40  \\
  \texttt{phi\_resolution} & 80 & 200 & 40 \\
  \texttt{theta\_resolution} & 80 & 100 & 80 \\
  \texttt{high\_resolution} & 160 & 200 & 160 \\
  \hline
\end{tabular}
\end{table}

\begin{figure} [ht!]
  \centerline{\includegraphics[width=0.5\textwidth]{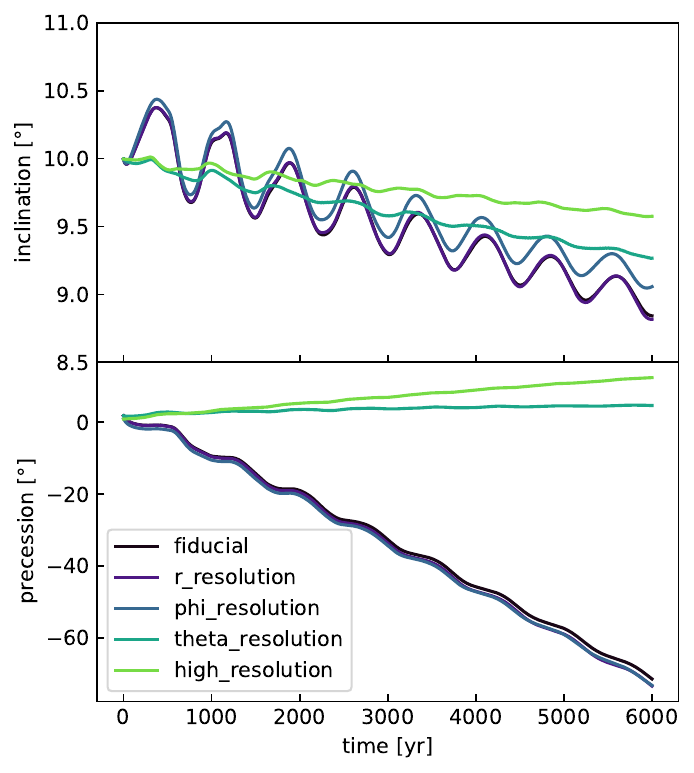}}
  \caption{\label{fig:10deg-resolution} Inclination (top panel) and precession angle (bottom panel) in simulations with varying resolutions in a disk tilted by 10$\degree$ with respect to the grid midplane.}
\end{figure}

Figure \ref{fig:10deg-resolution} shows the inclination and precession in the simulations with varying resolutions.
Looking at the inclination evolution, we find that the decrease of inclination is weakest for the simulation \texttt{high\_resolution}.
Comparing the three simulations with increase of resolution in one direction each, we find that the grid effects are especially sensitive for resolution in $\theta$-direction.
The inclination oscillations become weaker and the overall decrease is less steep.
Resolution in $r$-direction, however, shows no significant improvements compared to the fiducial \rev{set-up} from Section \ref{sec:tilteddisk}.

We find that resolution in $r$ and $\varphi$ do not influence the precession motion.
However, higher resolution in $\theta$ makes a large difference for the precession, as we have already suspected in Section \ref{sec:tilteddisk}.
In simulation \texttt{theta\_resolution}, there is almost no precession of the disk.
This indeed confirms our suspicion that the precession is mainly caused by friction of gas parcels that have to travel through several $\theta$-layers because of the tilt.
We note that simulation \texttt{high\_resolution} is slightly precessing in the other direction.
However, this precession is still small compared to the fiducial resolution.

Overall, we find that the grid effects get weaker for higher resolution, as expected.
The oscillation of the inclination and precession angle we investigated in Figure~\ref{fig:wobbles} becomes much weaker when increasing the $\theta$-resolution.
Higher resolution can severely increase the computation time.
Doubling the resolution in all three directions prolongs the computation by a factor of 16 ($2^3=8$ for the three directions and an additional factor of two due to the decreased time step).
However, the simulations we are investigating in this work are still feasible, because we can take the advantage of multi-node GPU parallelization.
For our purposes, we find the grid effects of simulation \texttt{theta\_resolution} to be sufficiently small.
We will therefore continue using this resolution for our investigations of a warped disk in Section~\ref{sec:warpeddisk}.

\subsection{Dependence on Viscosity} \label{sec:tilteddisk-viscosity}

To investigate the role of viscosity for the grid effects, we perform four further simulations varying the viscosity parameter $\alpha_\mathrm{t}$.
For these simulations, we use the initial low resolution (see Section~\ref{sec:tilteddisk}).
We keep the \rev{set-up} exactly the same except for the viscosity, which we set to $\alpha_\mathrm{t}=10^{-5}$, $\alpha_\mathrm{t}=10^{-4}$, $\alpha_\mathrm{t}=10^{-2}$, and $\alpha_\mathrm{t}=10^{-1}$.
The fiducial simulation we investigated in previous sections has a viscosity of $\alpha_\mathrm{t}=10^{-3}$.

\begin{figure} [ht!]
  \centerline{\includegraphics[width=0.5\textwidth]{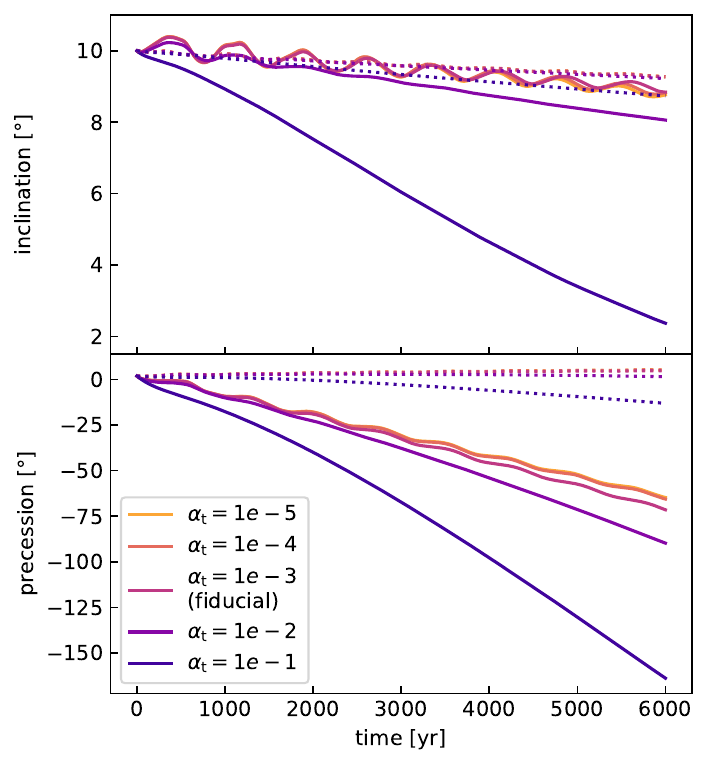}}
  \caption{\label{fig:10deg-viscosity} Same as Figure~\ref{fig:10deg-resolution}, but for simulations using varying viscosities. The disks are still tilted by $10\degree$ with respect to the grid midplane. \rev{The dotted lines show the same simulations with higher resolution in $\theta$-direction (corresponding to \texttt{theta\_resolution} in Table~\ref{tab:resolutions}).}}
\end{figure}

Figure~\ref{fig:10deg-viscosity} shows the evolution of the inclination and the disk precession in those simulations.
The simulations with $\alpha_\mathrm{t}=10^{-5}$, $\alpha_\mathrm{t}=10^{-4}$, and  $\alpha_\mathrm{t}=10^{-3}$ show a very similar behaviour with almost the same amount of decrease in inclination and precession angle.
\rev{For higher} viscosities, however, \rev{the grid effects seem to be much stronger.
We investigated the same simulations in higher resolution (corresponding to \texttt{theta\_resolution}) and find that the grid effects significantly decrease for all viscosities (dotted lines in Figure~\ref{fig:10deg-viscosity}). However, the simulation with the highest viscosity, $\alpha_\mathrm{t}=10^{-1}$, still shows slightly stronger numerical effects than the other simulations. }
This means that in order to accurately model higher viscosities, \rev{it is even more important to consider a higher resolution}.
In this work, we are mainly interested in viscosities of $\alpha_\mathrm{t}=10^{-3}$ and lower and therefore \rev{do not go into more detail here.}
It should, however, be kept in mind for future projects.

\subsection{Higher Inclination}

In order to study the grid effects for a different tilt inclination, we perform a simulation of a disk tilted by $30\degree$ with respect to the grid midplane.
For this, we decided to use a higher resolution, i.e. the same as in \texttt{simulation high\_resolution}.
Since our disk is more inclined, we need to extend the computation domain in $\theta$-direction.
In order to achieve the same resolution in $\theta$ for this case, we need more grid cells in $\theta$.
Table \ref{tab:30deg} shows the resolution comparison between the two simulations.
We keep the cell amounts in $r$ and $\varphi$ the same.

\setlength\extrarowheight{5pt}
\begin{table}[h]
\caption{Resolution Comparison Between Different Inclinations} \label{tab:30deg}
\begin{tabular}{ | m{2.7cm} | m{0.9cm}| m{0.75cm} | m{0.75cm} | m{1.4cm} |}
  \hline
  simulation name & $\theta_\mathrm{min}$ & $\theta_\mathrm{max}$ & cells in $\theta$ & degree per cell \\
  \hline
  \hline
  \texttt{high\_resolution} & -25.8$\degree$ & 25.8$\degree$ & 160 & 0.322$\degree$/cell \\
  \texttt{30deg} & -43.0$\degree$ & 43.0$\degree$ & 264 & 0.326$\degree$/cell \\
  \hline
\end{tabular}
\end{table}

\begin{figure} [ht!]
  \centerline{\includegraphics[width=0.5\textwidth]{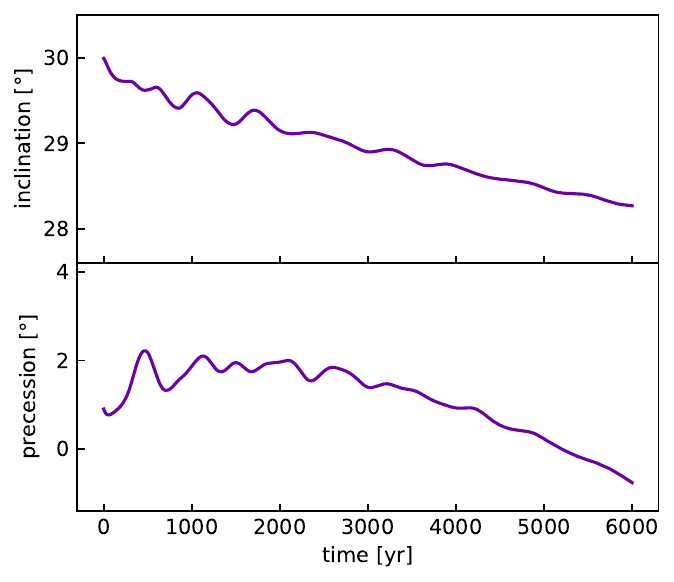}}
  \caption{\label{fig:30deg} Inclination (top panel) and precession (bottom panel) evolution of a disk tilted by 30$\degree$ with respect to the grid midplane.}
\end{figure}

Figure \ref{fig:30deg} shows that the inclination decreases by about two degrees over a time span of 6000 years.
\rev{We can roughly estimate the decay rate $\lambda = - 1/i \ \mathrm{d}i/\mathrm{d}t = - 2\degree/30\degree \ 1/(6000\,\mathrm{yr})$ to $\lambda = - 1.1 \cdot 10^{-5}\,\mathrm{yr^{-1}}$.
When comparing this to the simulation  \texttt{high\_resolution} with $\lambda = - 0.5\degree/10\degree \ 1/(6000\,\mathrm{yr}) = - 8.3 \cdot 10^{-6}\,\mathrm{yr}^{-1}$, we see that the decay rate for higher inclinations is higher.
We therefore expect the inclination decay not to be perfectly linear.}
For future projects investigating higher inclinations, we keep in mind that more tests should be made.
The precession of the disk varies a little, but the change stays within a few degrees.

All in all, we find that the grid effects can be negligible for a high enough resolution.
The resolution we find is still feasible for full three-dimensional hydrodynamic simulations.
We therefore conclude that it is possible to simulate disks with out-of-plane features using a grid-based method, when keeping in mind that grid effects might play a role under certain conditions.

\section{Warped Disk} \label{sec:warpeddisk}

To investigate the warp evolution in three-dimensional hydrodynamics, we set up an initially warped disk around a single central star.
Note that the shape of the warp is not physically motivated by any formation scenario, as we are interested in the evolution from that point on.
The initial warp is given in Equation~\ref{eq:incl-ini}.
Since we do not include a misaligned binary or companion or any other external component, the warp is not driven and will smooth out according to the warp evolution.
The initial maximum warp is $i_\mathrm{max}=10\degree$.
We locate the warp to $r_\mathrm{warp}=10.4\,\mathrm{au}$ with a width of $r_\mathrm{width}=2.6\,\mathrm{au}$.
For the resolution, we choose 80 cells in $r$, 100 cells in $\varphi$, and 134 cells in $\theta$ with an extend of $-43\degree<\theta<43\degree$, which keeps the cells-per-scaleheight ratio the same \rev{(i.e., 4.4~cps at $r=5.2\,\mathrm{au}$)} as in the simulation \texttt{theta\_resolution} in Section~\ref{sec:tilteddisk}.
We chose this to keep the computational time short so we can perform multiple simulations.
We performed one high-resolution simulation doubling the resolution in all three directions and find that the qualitative results are the same.
\rev{Our evaluation of this high resolution simulation can be found in Appendix~\ref{sec:appendix-highres}.}

\begin{figure} [ht!]
  \centerline{\includegraphics[width=0.5\textwidth]{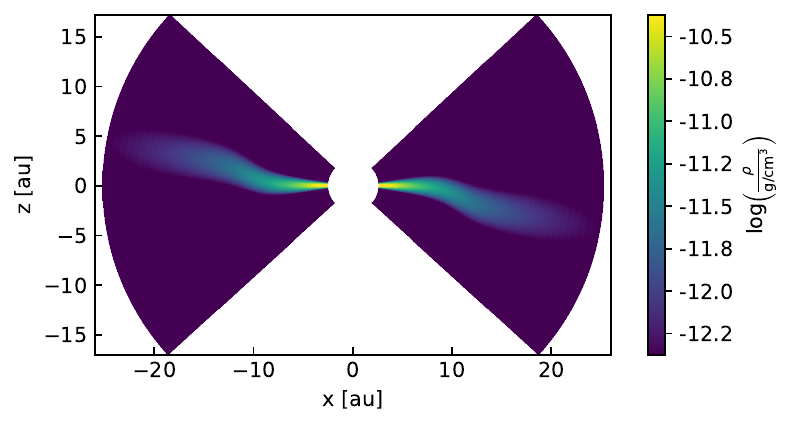}}
  \caption{\label{fig:warp-setup} \rev{Cross-section} of the initial warped disk along the x-axis.}
\end{figure}

Figure~\ref{fig:warp-setup} shows a cross-section of this initial \rev{set-up}.
Except for the warping of the disk, all parameters are the same as in our previous simulations (see Section \ref{sec:reference-case}).
We again choose a turbulence parameter of $\alpha_\mathrm{t}=10^{-3}$ and therefore expect wave-like evolution of the warp, as $h>\alpha_\mathrm{t}$.
In the following sections, we investigate the evolution of the inclination profile, as well as the precession profile (=twist) of the disk.
We expect the profiles to depend on the radius, and therefore do not look at the radially averaged inclination and precession angle.

\subsection{Inclination Profile}

We first investigate the evolution of the inclination profile in detail.

\begin{figure} [ht!]
  \centerline{\includegraphics[width=0.5\textwidth]{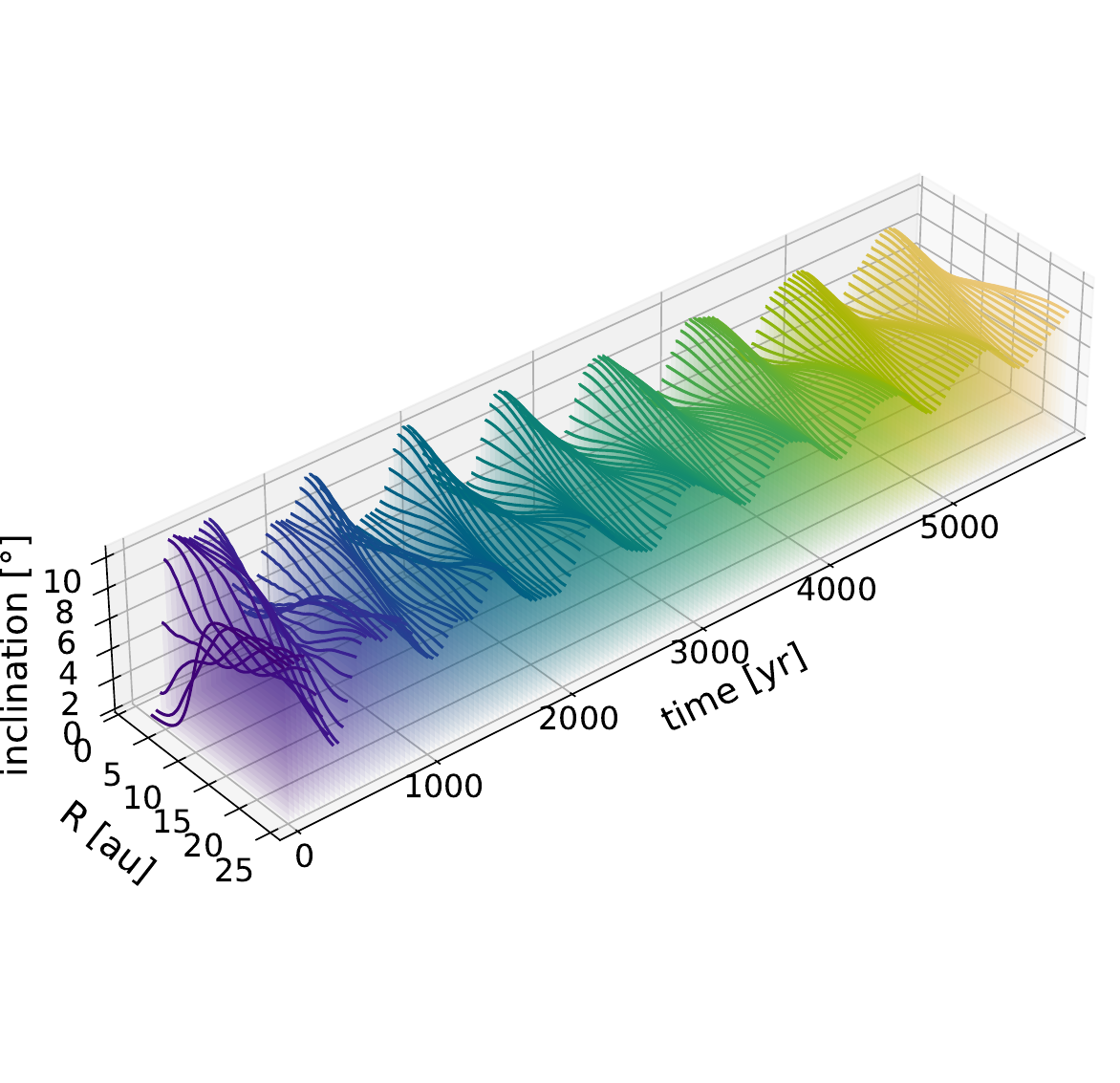}}
  \caption{\label{fig:warp-incl} Evolution of the inclination profile in the warped disk simulation. The color simply highlights the time.}
\end{figure}

In Figure \ref{fig:warp-incl}, we \rev{show the} evolution of the inclination profile.
\rev{In this figure, the wave-like behavior is evident and is visible as a standing wave of the global bending modes superimposed on a global mean tilt of the disk. This mean tilt of the disk does not visibly decay, although we expect it to decay slowly due to numerical effects.}
The warp amplitude is dampened over time.
During the first 6000 years (500 orbits at $R_0$), the disk does not reach the planar state.

\begin{figure} [ht!]
  \centerline{\includegraphics[width=0.5\textwidth]{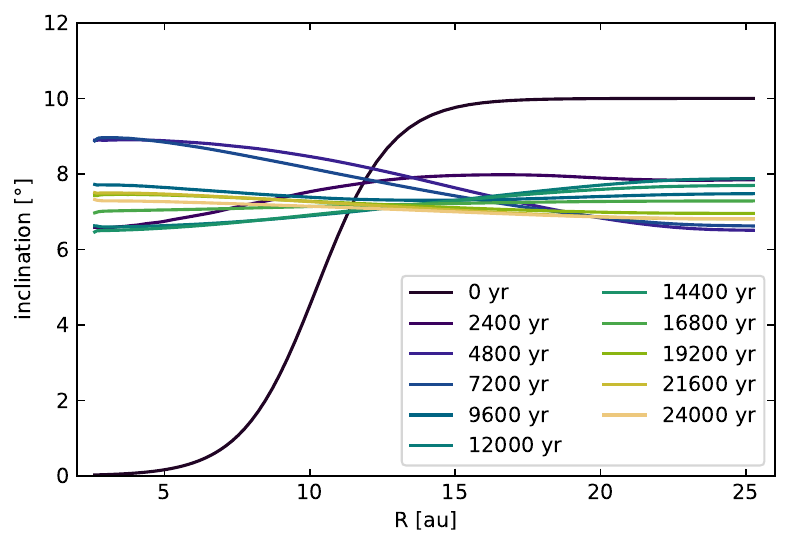}}
  \caption{\label{fig:warp-incl-long} Long-term evolution of the inclination profile in the warped disk simulation.}
\end{figure}

To investigate the damping of the warp, we continue running the simulation up to $2.4 \cdot 10^4\,\mathrm{yr}$ (2000 orbits at $R_0$).
Figure~\ref{fig:warp-incl-long} shows this long-term evolution.
The warp is dampened so that we end up with a nearly planar, \rev{however, spatially tilted} disk.

\begin{figure} [ht!]
  \centerline{\includegraphics[width=0.5\textwidth]{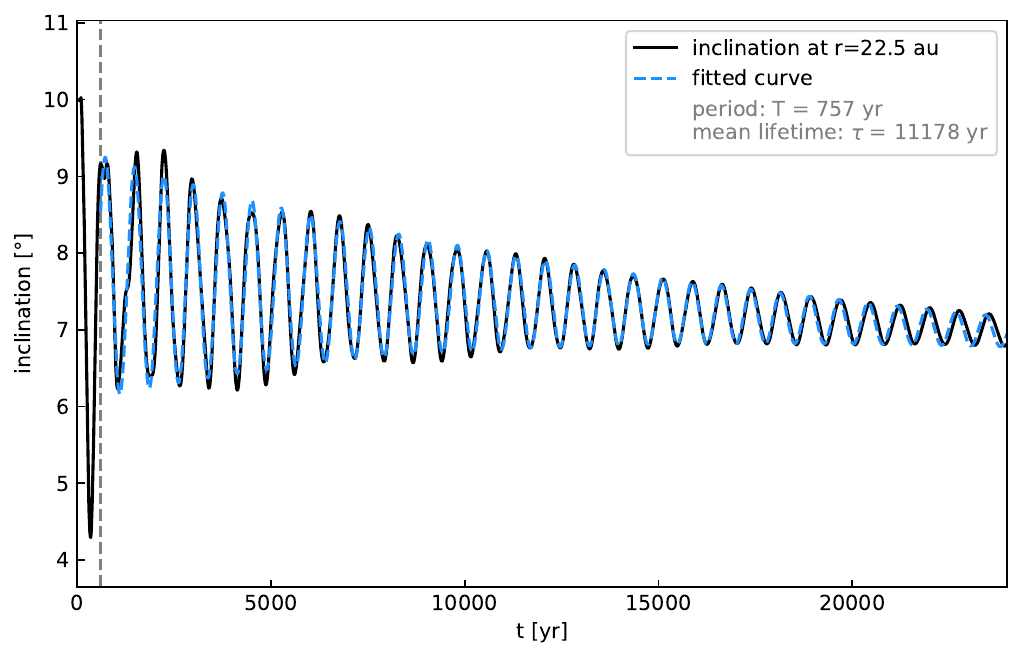}}
  \caption{\label{fig:incl-decay} \rev{Decay of inclination in the warped disk simulation. We chose to evaluate the inclination close to the outer edge at a radius of $22.5\,\mathrm{au}$. The black line shows the simulation data, the blue dashed line the fit. The grey dashed line indicates the starting point of the fit.}}
\end{figure}

\rev{In Figure~\ref{fig:incl-decay}, we investigate the warp wave period and inclination damping in more detail. We choose a radius close to the outer edge of the disk at $r = 22.5\,\mathrm{au}$ and fit the inclination evolution at this point with an exponentially decaying cosine function of the form
\begin{equation}\label{eq:fit}
f(t) = A \cos(\omega t - b) \cdot \exp(-\lambda t) + c + d \cdot t,
\end{equation}
with amplitude parameter $A$, frequency $\omega$, and decay rate $\lambda$. We allow for a phase shift $b$ in the cosine function, an offset $c$, and a linear damping in overall inclination with $d$. \newline
Our fit gives a wave frequency of $\omega = 0.008\,\mathrm{yr}^{-1}$, which leads to a warp wave period of
\begin{equation}
P = \frac{2 \pi}{\omega} = 757\,\mathrm{yr}.
\end{equation}
We can make an estimate of the period from linear theory in the following way. The wave speed of the warp wave is approximately half the sound speed in the disk $v_\mathrm{w} = c_\mathrm{s} / 2$ \citep{LubowOgilvie2000}. Because we are investigating a globally isothermal set-up for now, the sound speed is $c_\mathrm{s} = 653\,\mathrm{m/s}$ everywhere in the disk. We find the warp wave to behave as a standing wave with a wave length of twice our radial disk regime $\lambda_\mathrm{w} = 2 \Delta r= 44.8\,\mathrm{au}$. The estimated period then is
\begin{equation}
P_\mathrm{linear\ theory} = \frac{\lambda_\mathrm{w}}{v_\mathrm{w}} = \frac{2 \Delta r}{2 c_\mathrm{s}} = 657\,\mathrm{yr}.
\end{equation}
Note that this is the same as the sound-crossing time $t_\mathrm{s} = \Delta r / c_\mathrm{s}$.
This means that the period in our simulation is in agreement with linear theory.
}

\rev{In our fit, we find a decay rate of $\lambda = 8.9 \cdot 10^{-5}\,\mathrm{yr}^{-1}$, which results in a mean lifetime of the warp of $\tau = 1/\lambda = 1.1\cdot 10^4\,\mathrm{yr}$. Linear theory gives a rough estimate of the warp damping timescale with $\tau_\mathrm{linear\ theory} = 1/(\alpha \Omega)$ with $\Omega$ as Keplerian frequency. \citep[see e.g.][]{Nixon2016}. If we choose the Keplerian frequency at the outer edge of our disk, this results in $\tau_\mathrm{linear\ theory} = 2.1\cdot 10^4\,\mathrm{yr}$, which agrees within a factor of 2 with the findings in our simulation.}

\rev{From our fit, we also get the mean tilt of the disk ($c$ in Equation~\ref{eq:fit}) at our chosen radius, which is $7.7\degree$, and the damping of inclination $d$, which we find to be $-3 \cdot 10^{-5}\,\mathrm{yr}^{-1}$, which gives a total inclination loss over the simulated $2.4\cdot 10^4\,\mathrm{yr}$ of $0.74\degree.$ This loss of mean inclination is a numerical effect as also found in Section~\ref{sec:tilteddisk}.
Interesting to note is also the amplitude from the fit with $A = 1.6\degree$.}

We now compare the three-dimensional hydrodynamic simulations to a one-dimensional model of warped disks.
For the 1D model, we split the disk into concentric annuli and assign each annulus an angular momentum vector that determines the orbital plane of the annulus.
This way, we have a one-dimensional parametrization for a three-dimensional object.
The angular momentum vector can be updated using Equations \ref{eq:mass-conservation}--\ref{eq:internal-torque}.
We set up the parameters in the 1D model the same parameters as in the 3D model.
For this, we use our 1D warped disk evolution code \texttt{dwarpy} (Kimmig et al. in prep).

\begin{figure} [ht!]
  \centerline{\includegraphics[width=0.5\textwidth]{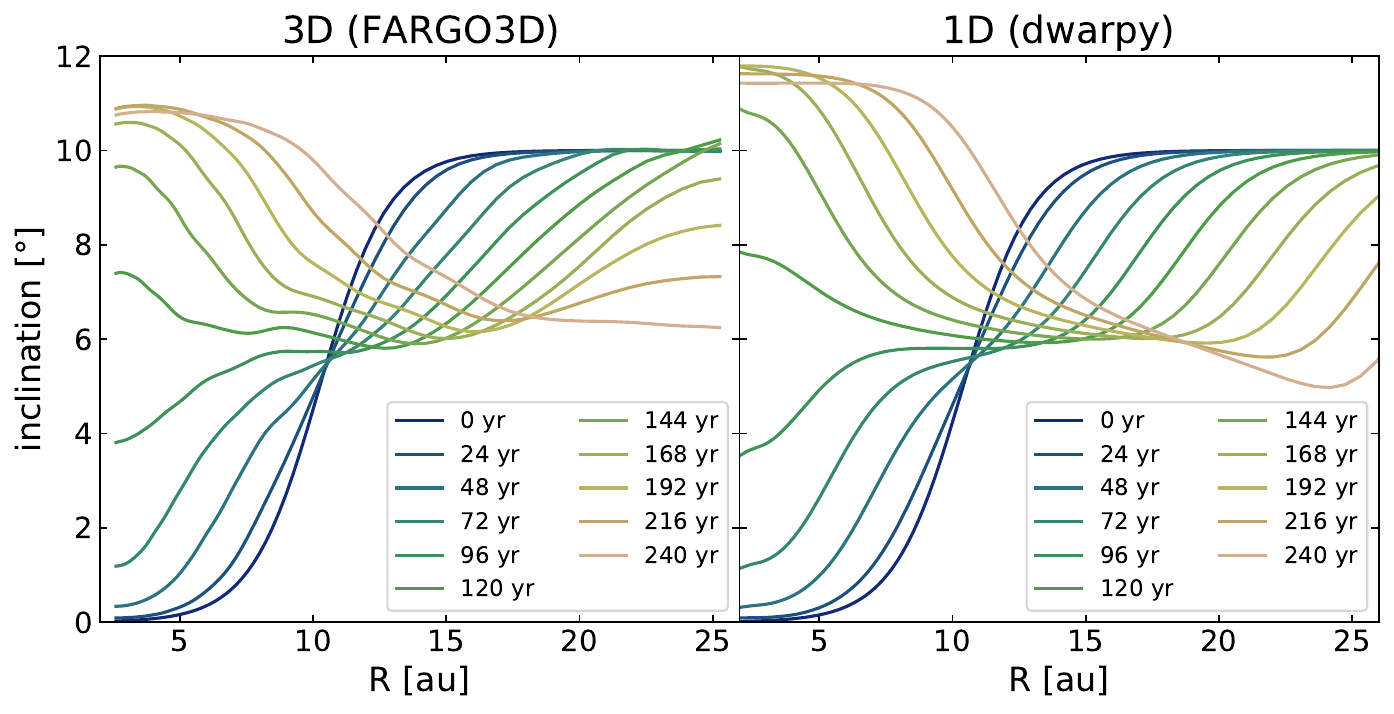}}
  \caption{\label{fig:fargo-dwarpy-comparison} Inclination evolution of the warped disk in comparison between the 3D \rev{set-up} in FARGO3D (left panel) and the 1D \rev{set-up} in dwarpy (right panel). For comparability, we plot until 240 years of simulated time. \rev{Note that the computational domain of the 1D simulation is well beyond the outer edge of the figure.}
  }
\end{figure}

\begin{figure} [ht!]
  \centerline{\includegraphics[width=0.5\textwidth]{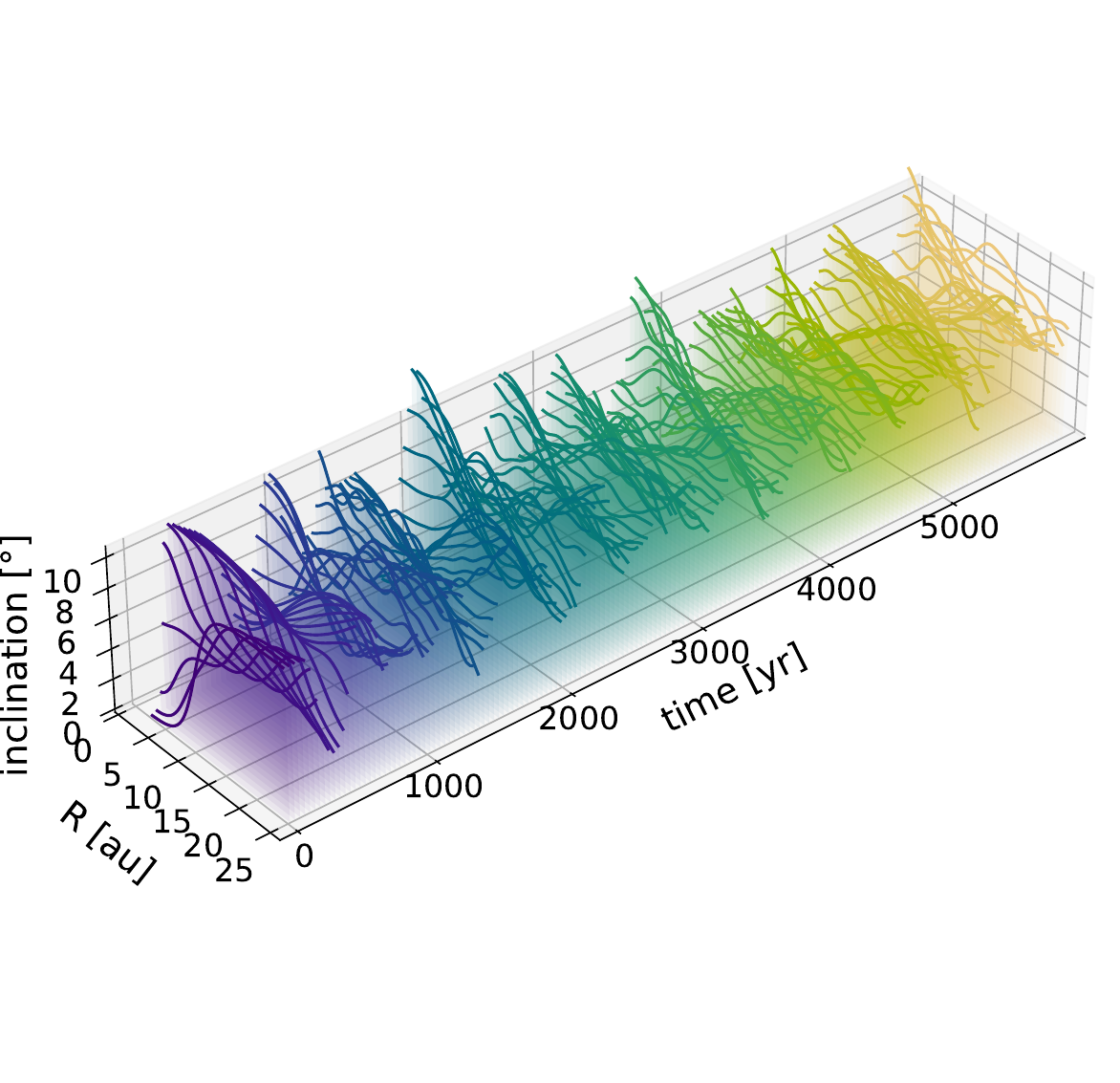}}
  \caption{\label{fig:warp-incl-dwarpy} Like Figure~\ref{fig:warp-incl}, but for the 1D model with \texttt{dwarpy}.}
\end{figure}

We find in Figure \ref{fig:fargo-dwarpy-comparison} a good agreement between three-dimensional and one-dimensional simulations on short time scales.
On longer time scales, however, the one-dimensional model shows a more \rev{complicated} behaviour than the 3D simulation, as shown in Figure~\ref{fig:warp-incl-dwarpy}.
This might be due to boundary conditions, as we used rigid boundaries in the 3D simulation and open boundaries in the 1D model, as \texttt{dwarpy} is primarily tested with these conditions. There, we chose open conditions in order to avoid a torque from the boundaries acting on the disk.
However, the wave-like nature of the inclination evolution is still apparent in the one-dimensional model.

\subsection{Twisting of the Disk}

In our three-dimensional simulation, we notice a precession motion within the disk, which does not occur in the one-dimensional model for the wave-like regime.
The three-dimensional disk shows a differential precession, meaning that the disk precesses differently depending on the radius, which we call a \textit{twist}.
In this section, we \rev{investigate} the twist in \rev{more} detail.
We aim to find out whether the twist is caused by numerical or physical effects.


\rev{At first}, we look at the total angular momentum of the disk.
We find that its absolute value decreases about $0.5\,\%$ in the time span of $6000\,\mathrm{yr}$ and $2\,\%$ in the time span of $2.4\cdot 10^4\,\mathrm{yr}$.
Evaluating the total angular momentum of the planar reference case from Section \ref{sec:reference-case}, we find that the planar disk's absolute angular momentum also decreases by about $0.5\,\%$ in the time span of $6000\,\mathrm{yr}$.

\begin{figure} [ht!]
  \centerline{\includegraphics[width=0.5\textwidth]{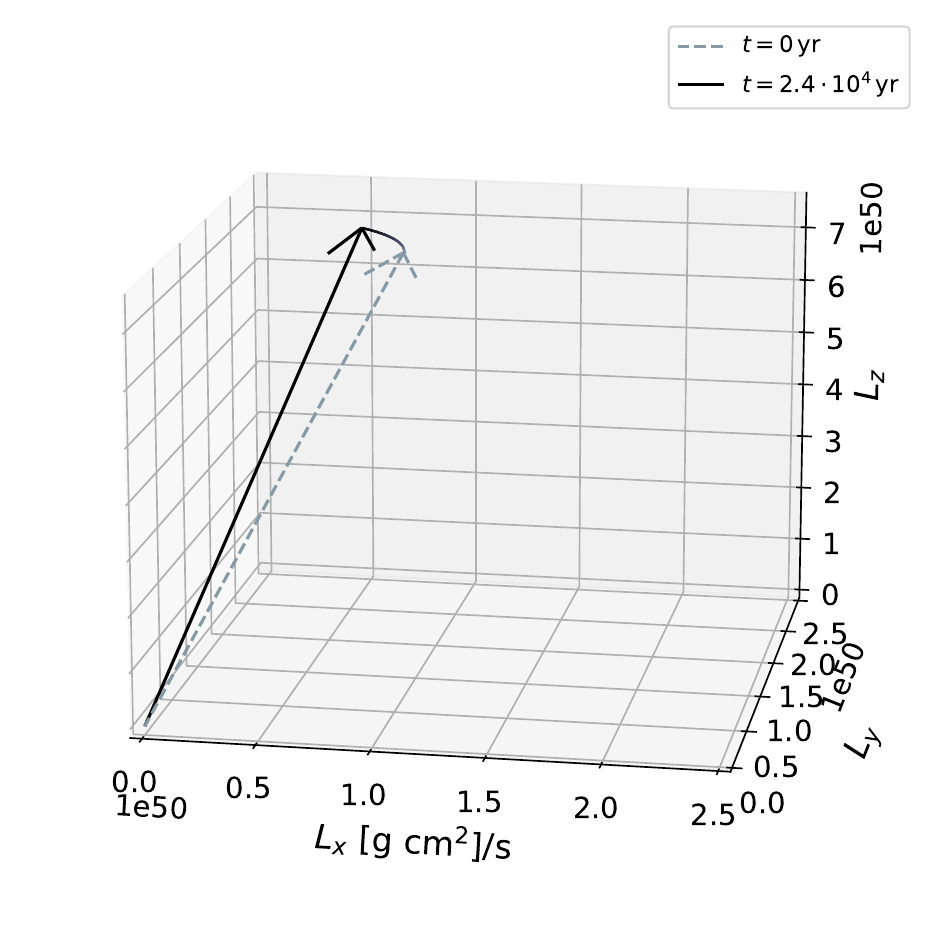}}
  \caption{\label{fig:tot-angmom-direction} Direction of the total angular momentum vector in the beginning of the simulation $t=0\,\mathrm{yr}$ (grey dashed line) and in the end $t = 2.4 \cdot 10^4\,\mathrm{yr}$ (solid black line). The line connecting the two arrows indicates the time evolution with darker color the later the time.
  }
\end{figure}

Figure \ref{fig:tot-angmom-direction} shows that the direction of the total angular momentum vector slightly changes over a time span of $2.4\cdot 10^4\,\mathrm{yr}$, which are $2000$ orbits at $R_0$.
The angle between the vectors at the beginning and the end of the simulation is $4.6\degree$.
We suspect that the change in direction occurs due to numerical effects.
\rev{While FARGO3D is programmed such that it conserves azimuthal angular momentum to machine precision, this mainly applies to components perfectly aligned with the grid geometry.
We suspect the change in angular momentum we observe in Figure~\ref{fig:tot-angmom-direction} to be caused by the fact that the numerical scheme does not account for tilted motions. }
On a time scale of $6000\,\mathrm{yr}$ (inclination evolution in this time span plotted in Figure~\ref{fig:warp-incl}), on which we find a significant twisting of the disk, we find the angular momentum vector to change on an angle of only $1.2\degree$.
Notably, the total angular momentum vector does not show any oscillatory behaviour.
\rev{In spite of the slight changes in angular momentum, we can argue that the changes are small and therefore,} the total angular momentum is reasonably well conserved, both in direction and in absolute value.

\begin{figure} [ht!]
  \centerline{\includegraphics[width=0.5\textwidth]{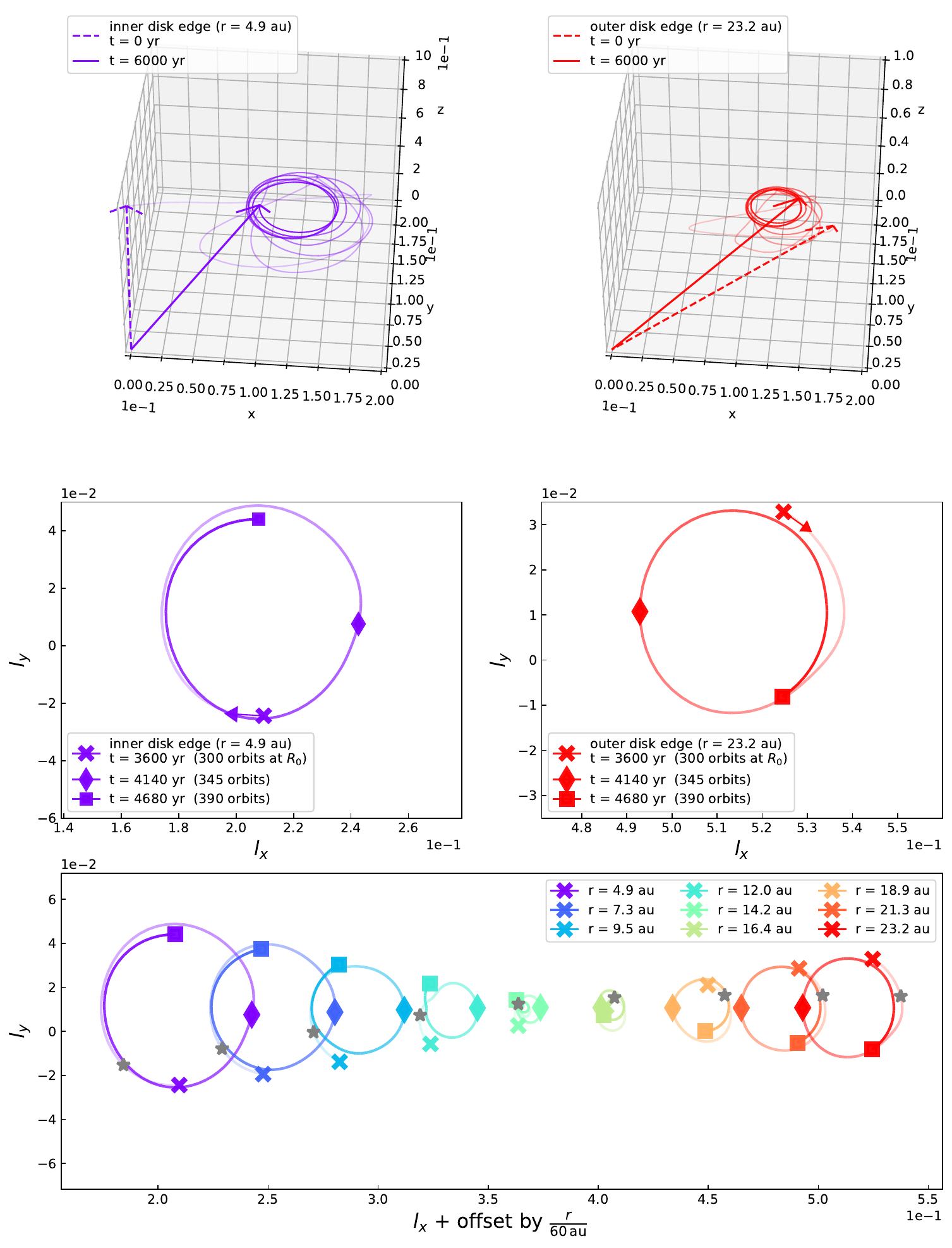}}
  \caption{\label{fig:precession-radii} \rev{Twist motion in the warped disk simulation.} \textbf{Top row:} Unit angular momentum vectors close to the inner (left panel) and outer (right panel) edge of the disk for the first $t~=~6000\,\mathrm{yr}$. Similar to Figure~\ref{fig:tot-angmom-direction}.
  \textbf{Middle row:} $x$-$y$-components of the unit angular momentum vector at the same radii as in the top row, but only for the time period between 300-390 orbits at $R_0$, chosen to roughly show 1.5 precession periods. The direction of the precession is indicated by the arrows. x-markers show the beginning of the plotted time, squares the end. Diamonds indicate the middle of this time period (45 orbits after the x-markers).
  \textbf{Bottom row:} as middle row, but for various radii, plotted with an offset of $r/(60\,\mathrm{au})$ in $y$-direction to disentangle the twist at different radii.
   The stars show the $l_x$-$l_y$-components at a snapshot at $t~=~3696\,\mathrm{yr}$ (308 orbits at $R_0$), which shows a twisted state in the time evolution.
  }
\end{figure}

We \rev{now} plot the twist of the disk by showing the evolution of the unit angular momentum vectors at different radii in Figure~\ref{fig:precession-radii}.
The angular momentum vectors are extracted from the three-dimensional simulation using Equation~\ref{eq:angmom-simulation}.


A twist occurs when the unit angular momentum vectors point in different directions.
In the top row of Figure \ref{fig:precession-radii}, we show the time evolution of the vectors close to the inner disk edge (left) and close to the outer edge (right).
We intentionally do not choose the exact disk edges to avoid potential boundary effects.
The disk quickly twists. After some time, the vectors precess in almost perfect circles.

To closer investigate this motion, we choose a point in time (300 orbits at $R_0$) at which the circles are already developed and plot in the middle row one and a half periods from that point on.
Here, we only plot the $x$- and $y$-component of the unit angular momentum vector $\vec{l}_r$.
We find the period to be roughly 60 orbits.
We see that both the inner and the outer part precess clockwise. However, because the total angular momentum is conserved in direction (as seen \rev{in Figure~\ref{fig:tot-angmom-direction}}), the directions counteract, which means that whenever the inner part of the disk points in positive $x$- and/or $y$-direction, the outer part points correspondingly in negative $x$- and/or $y$-direction.
To see this, we indicate a few corresponding times with the symbols in the plot.

In the bottom panel of Figure~\ref{fig:precession-radii}, we then plot more radial points in the disk, each radius in a different color. Note that we offset the twist for each radius by a factor of $r/(60\,\mathrm{au})$ in order to disentangle the circles.
We chose a linear spacing for the radial evaluation points.
At all radii, the disk precesses clockwise, and the counteracting of the directions can be seen in this plot in more detail.
One part of the disk always counteracts the precession of another part, so that the total angular momentum is conserved.
A clearly twisted state of the disk occurs for example in the snapshot at 308~orbits, which is illustrated with the grey stars.

We note that in the long-term evolution of the system, the precession is dampened.
This happens on the same time scale as the inclination of the warp is dampened.
Additionally, the precession period of the vectors coincides with the inclination wave period.

To explore the origin of the disk's twist — whether it is a numerical or physical effect — we conducted a series of tests, \rev{which can be found in Appendix~\ref{sec:appendix-twist}, and we briefly summarize:}
\rev{The fact that the total angular momentum does not exhibit any oscillatory motion} led us to rule out the inner and outer grid boundary as a cause for the twist.
To investigate further, we performed a new simulation, pushing the outer boundary far away from the physical grid edge by applying an exponential cut-off to the disk density.
In this scenario, the outer boundary is unlikely to influence the physical outer edge of the disk.
However, even in this case, the twist persisted. \newline
Furthermore, we performed one-dimensional simulations to explore if the twist might be triggered \rev{by the initial conditions} in the 3D simulation.
However, the 1D model failed to reproduce the twist observed in the 3D model.
This strongly suggests that the twist is indeed a three-dimensional phenomenon. \rev{Interestingly, a similar effect can be seen in SPH simulations of warped disks (R. Martin, priv. comm.).} \newline
\rev{While} we can not be completely sure at this point, our tests indicate that the twist likely arises from a physical cause rather than being a numerical artefact.
We suspect that specific terms neglected in the linearisation process in deriving the one-dimensional equations play a role in why the 1D model can not reproduce the twist.

\subsection{\rev{Investigating a Differently Flared Disk}} \label{sec:fl25}

\begin{figure} [ht!]
  \centerline{\includegraphics[width=0.5\textwidth]{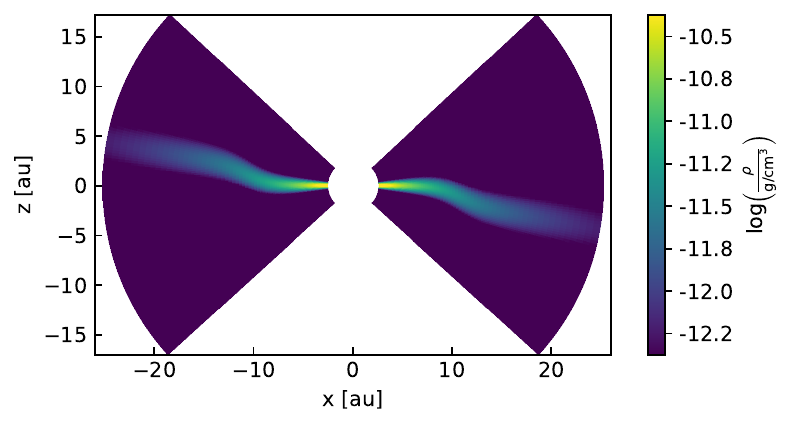}}
  \caption{\label{fig:warp-setup-fl25} \rev{Cross-section along the x-axis of the locally isothermal warped disk.}}
\end{figure}

\rev{In order to relax the assumption of a globally isothermal disk, we additionally run a simulation with a locally isothermal disk using a flaring index of $i_\mathrm{fl} = 0.25$ (see Equation~\ref{eq:aspectratio}) and setting the temperature structure accordingly. We keep all other parameters the same. A cross-section of the initial set-up is shown in Figure~\ref{fig:warp-setup-fl25}.
}

\rev{Investigating the inclination evolution, we find the expected wave-like behavior, as shown in Figure~\ref{fig:warp-setup-fl25}.
}

\begin{figure} [ht!]
  \centerline{\includegraphics[width=0.5\textwidth]{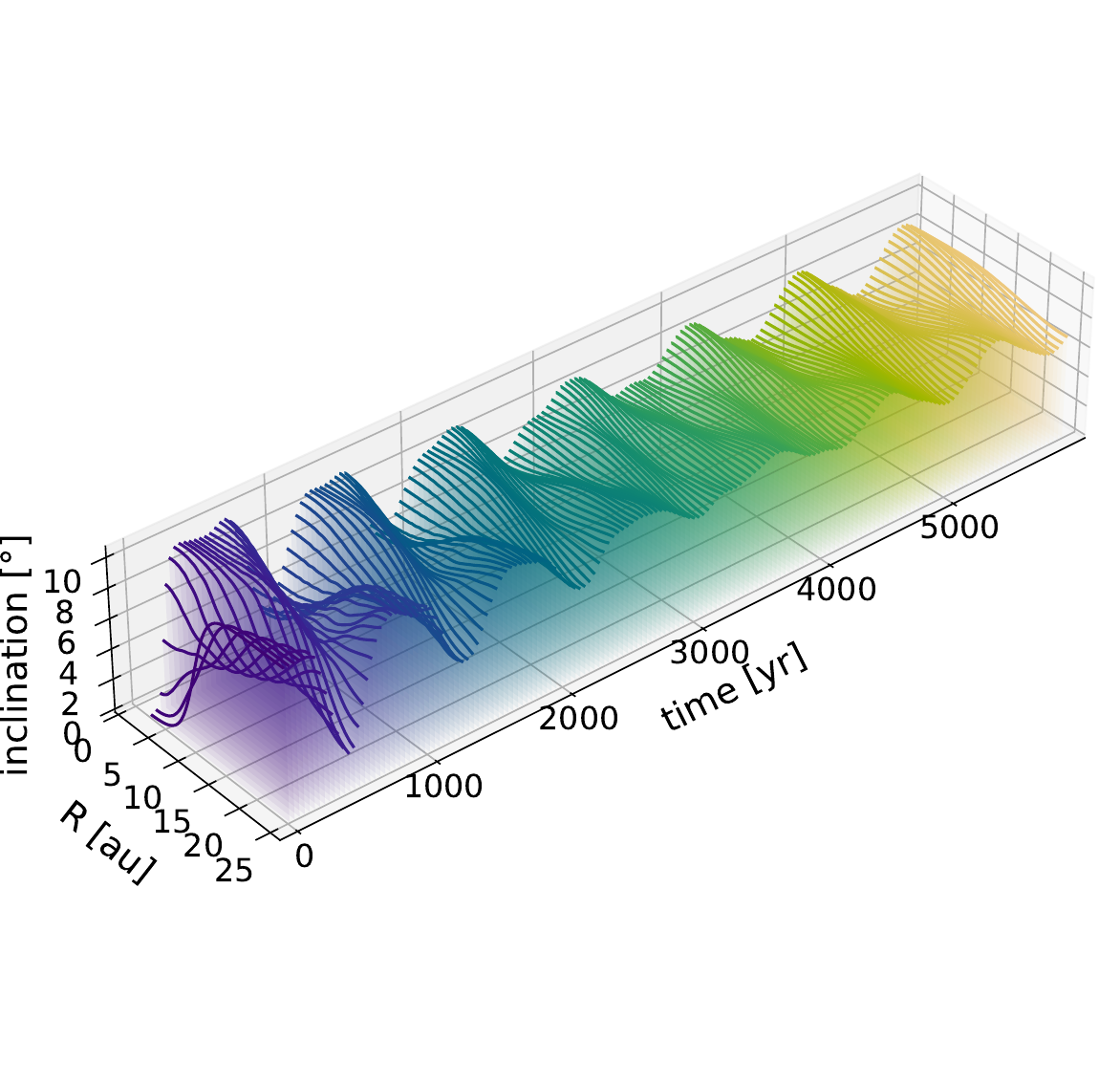}}
  \caption{\label{fig:warp-incl-fl25} \rev{Evolution of the inclination profile in the locally isothermal warped disk. Like Figure~\ref{fig:warp-incl}.}}
\end{figure}

\begin{figure} [ht!]
  \centerline{\includegraphics[width=0.5\textwidth]{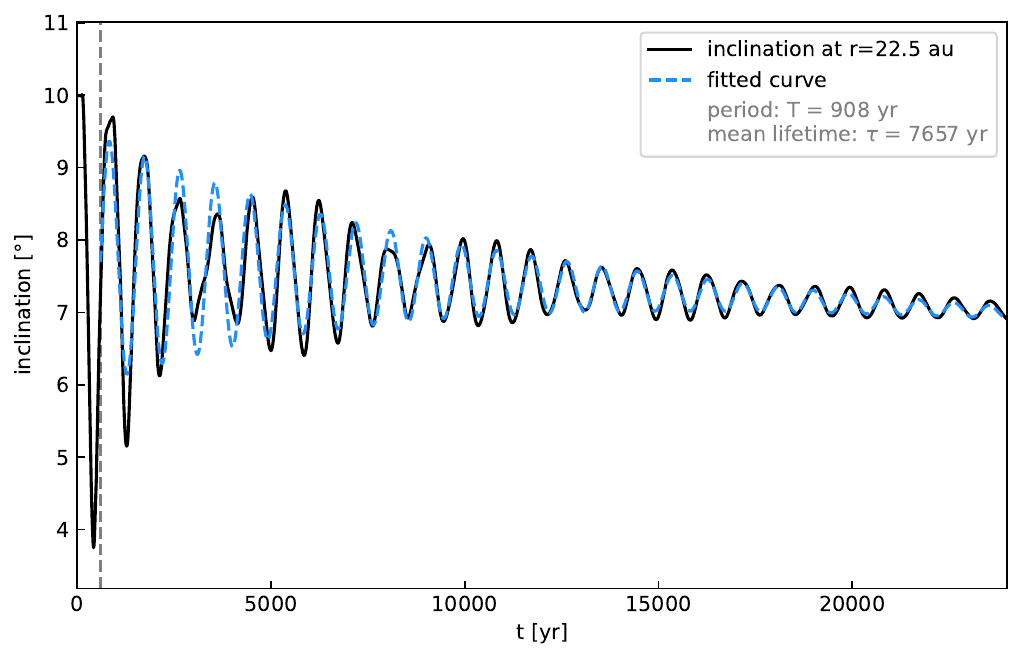}}
  \caption{\label{fig:incl-decay-fl25} \rev{Like Figure~\ref{fig:incl-decay}, but for the locally isothermal disk.}}
\end{figure}

\rev{The overall behavior is similar to the globally isothermal disk.
However, there seems to be an amplitude modulation with a period of about 4000\,yr, which can also be seen when we look at the inclination evolution at a specific radius (we choose $r=22.5\,\mathrm{au}$) as shown in Figure~\ref{fig:incl-decay-fl25} (black line).
This could be either a numerical or a physical effect, which would require further investigation. This would go beyond the scope of this work, thus we simply mention this observation at this point.
}

\rev{In Figure~\ref{fig:incl-decay-fl25}, we investigate the inclination decay rate according to Equation~\ref{eq:fit} and find a decay rate of $\lambda~=~1.3~\cdot~10^{-4}\,\mathrm{yr^{-1}}$, which gives a mean lifetime of $\tau~=~1/\lambda =~7.7\cdot 10^3\,\mathrm{yr}$.
This is slightly shorter than the fitted mean lifetime of the globally isothermal disk.
For reference, linear theory gives an estimation of $\tau_\mathrm{linear\ theory} = 1/(\alpha \Omega) = 2\cdot 10^4\,\mathrm{yr}$, which does not depend on the disk's vertical structure.
In future work, the influence of the vertical structure on the warp lifetime could be investigated in more detail.
}

\rev{In Appendix~\ref{sec:appendix-twist}, we investigate the twist of the locally isothermal disk. We find that it still occurs, but the twist behavior appears to be more complicated than in the globally isothermal case.}

\subsection{\rev{Different Viscosities}}

We further test the dependence of the \rev{warp and} twist on viscosity and perform two additional simulations, using the same initial \rev{set-up} but varying the original $\alpha_\mathrm{t}~=~10^{-3}$ to lower ($\alpha_\mathrm{t}~=~10^{-5}$) and higher viscosity ($\alpha_\mathrm{t}~=~10^{-2}$).

\begin{figure} [ht!]
  \centerline{\includegraphics[width=0.5\textwidth]{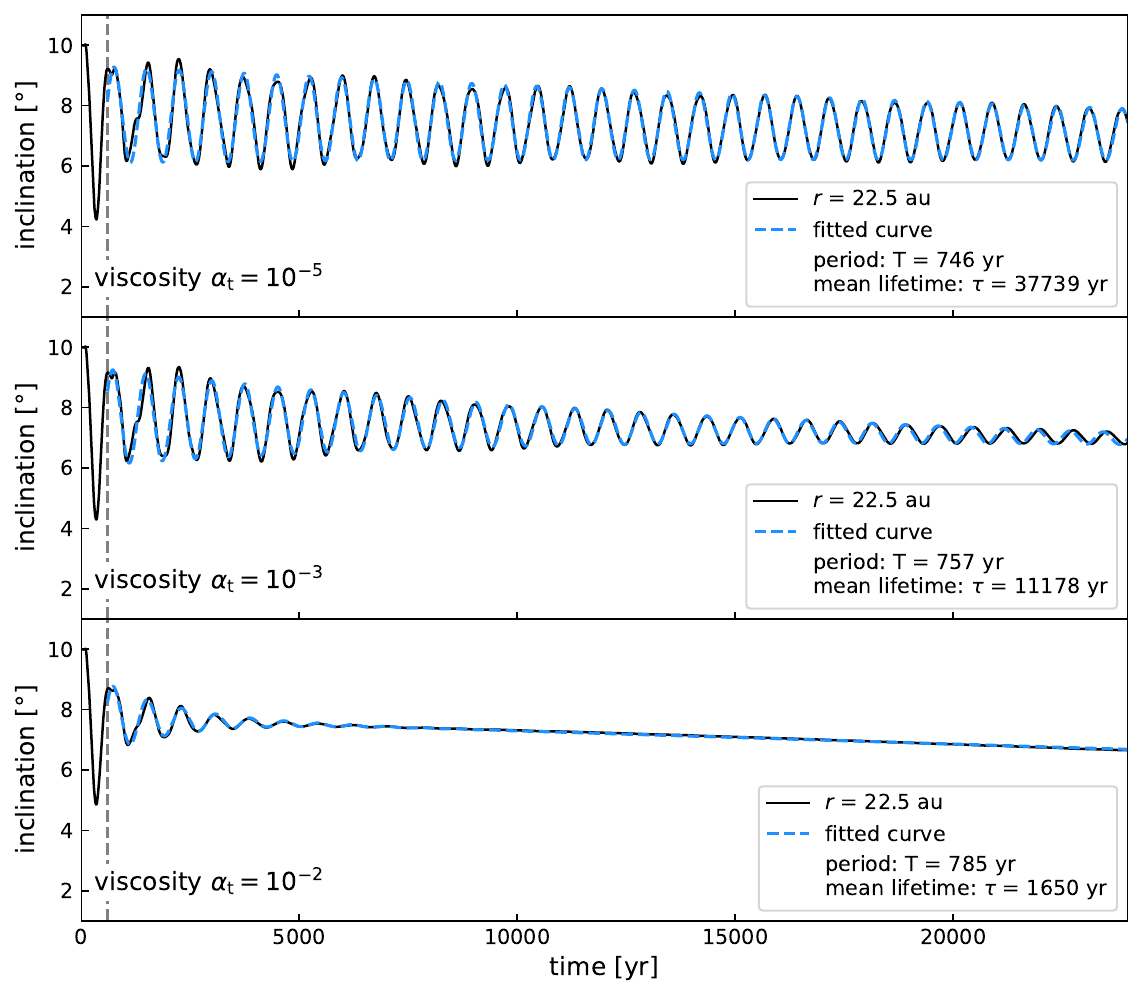}}
  \caption{\label{fig:incl-viscosities} \rev{Inclination evolution at radius $r=22.5\,\mathrm{au}$ (black solid lines) in simulations with different disk viscosities. The blue dashed lines indicate the fit according to Equation~\ref{eq:fit}. The middle panel corresponds to the fiducial simulation and is the same as Figure~\ref{fig:incl-decay}.
}  }
\end{figure}

\rev{We plot the inclination evolution at a radius $r=22.5\,\mathrm{au}$ (close to the outer edge) of the simulations with varying viscosities in Figure~\ref{fig:incl-viscosities} to investigate the warp decay rates.
We find that for a ten times higher viscosity (bottom panel), the warp is dampened roughly ten times faster than in our fiducial simulation, as expected by linear theory (recall $\tau_\mathrm{linear\ theory} = 1/(\alpha \Omega)$).
For the low viscosity simulation (top panel), the mean lifetime is longer than the fiducial simulation, as expected. However, instead of the factor 100 predicted by linear theory, we find a factor of roughly 4.
This could mean that the simulation is influenced by numerical viscosity and grid effects, or that additional effects play a role in dampening the warp that are not taken into account in the derivation of the time scale estimation.
Investigating the reason here in detail would go beyond the scope of this work.
}

\begin{figure} [ht!]
  \centerline{\includegraphics[width=0.5\textwidth]{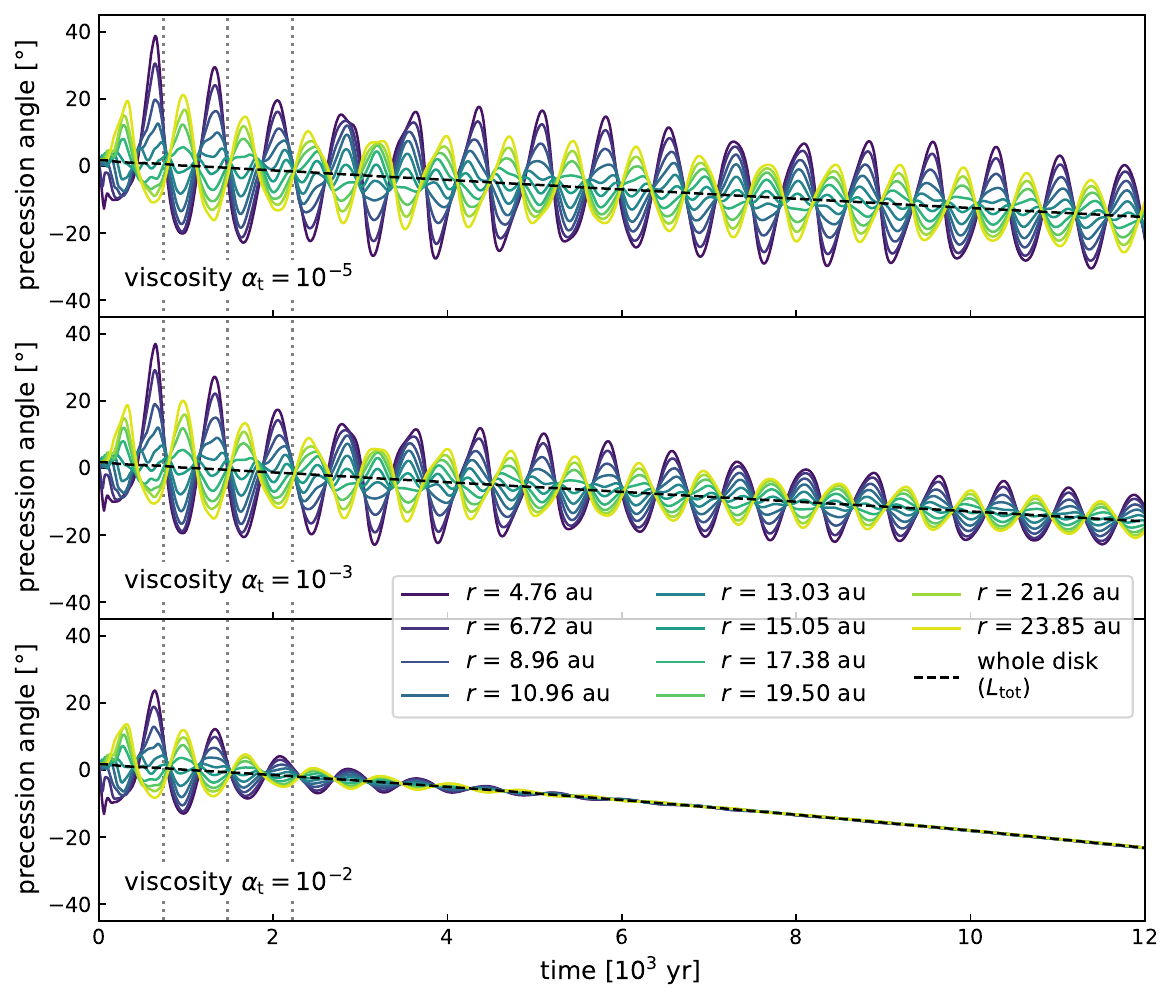}}
  \caption{\label{fig:prec-viscosities} Precession angles for different radii (\rev{colored} lines) in simulations with different disk viscosities. The black dashed line in each panel shows the precession angle of the angular momentum vector $\vec{L}_\mathrm{tot}$ of the whole disk. The grey dotted lines indicate precession periods of roughly $740\,\mathrm{yr}$.
  }
\end{figure}

In Figure~\ref{fig:prec-viscosities}, we plot the precession angle over time for the three different viscosities.
Recall that we define the precession angle as the clockwise angle between $x$-axis and angular momentum vector projected onto the $x$-$y$-plane, giving an initial precession angle of $0\degree$ at all radii.
We plot the precession angle at different radii (coloured lines).
As a reference, we additionally plot the precession angle of the total angular momentum vector (black dashed line) and note that it does not oscillate.

In all cases, we find that the total angular momentum vector changes direction in time, which means that the disk overall precesses.
This overall precession is more pronounced in the high viscosity case, $\alpha = 10^{-2}$.
This is not a physical effect, but a result from grid effects, as explored in Section \ref{sec:tilteddisk-viscosity}.
A higher resolution would help keep the direction of the angular momentum conserved for a longer time period.
However, since we are mainly interested in the differential precession, i.e. the twist, we do not perform additional simulations in higher resolution at this point.

In all cases, we find a twist, which is dampened much faster for high viscosity.
The precession period is similar for all viscosities, as indicated with the grey dotted lines in Figure~\ref{fig:prec-viscosities}, and roughly equals $740\,\mathrm{yr}$.

Despite that parts of the disk clearly precess, the total angular momentum direction \rev{does} not oscillate.
We take this as \rev{another} indication that the \rev{twisting} behaviour is physical, although the reason for \rev{it} is not clear yet.

\section{Internal Dynamics of the Warped Disk and the Origin of the Torques} \label{sec:sloshing}

\subsection{Sloshing Motions and Internal Torques}

The internal torque responsible for the propagation of a warp through the disk
is primarily due to resonant horizontal internal gas motions
\citep{PapaloizouPringle1983, PapaloizouLin1995, LubowOgilvie2000}. We refer to
these as ``sloshing motions'' because they are caused by sideways pressure
gradients due to the warp \citep[see Figure~5 of ][]{OgilvieLatter2013a}, and
behave very similar to a layer of water on a tilted tray. For any snapshot of a
3D warped disk model, the internal torque can be calculated as shown in
Appendix~\ref{sec-internal-torques}.

A description of how these internal sloshing motions are driven, how they
behave, and how they give rise to an internal torque, was given in DKZ22. There, the authors used a variant
of the local shearing box approach of \citet{OgilvieLatter2013a}. After
employing two linearisation steps, they derived the solutions for the sloshing
motions, and from those obtained a generalized set of one-dimensional warped disk
equations. With our three-dimensional model, we can investigate how good these approximate
solutions for the sloshing motions are, as we can directly analyse the motions
of the gas inside the 3D disk and compare them to the predictions of DKZ22.

Of course, we are not the first to compare 3D warped disk models to 1D models.
For instance, \citet{Nelson1999} performed 3D Smoothed Particle
Hydrodynamics simulations of warped disks in both the wavelike and diffusive
regime, and compared the results to 1D models. They also see the sloshing
motions in their 3D models. In our FARGO3D model in spherical
coordinates, we have the advantage compared to SPH models that we can study the disk
many scale heights above the midplane, where the densities are orders of
magnitude lower than at the midplane. While these regions do not contribute
significantly to the internal torque, they may contribute to observations
of the disk.

\subsection{Sloshing and Breathing Motions in the 3D Model}
To compare the internal dynamics of our 3D warped disk model with the local
shearing box approach of DKZ22, we choose a single annulus of the disk,
i.e., we choose a radius $r$ at which we perform the comparison. In DKZ22, the
global disk is assumed to be oriented such that the $\vec{l}$ vector (the unit
vector perpendicular to the disk annulus) at this radius points into the global
positive $z$-direction (see Figure~2 of DKZ22). Furthermore, the disk is oriented
such that the local warp vector $\vec{\psi}=\mathrm{d} \vec{l}/ \mathrm{d}\ln r$ points in the
positive $x$-direction. To compare our 3D disk model, at radius $r$, with the
local shearing box description of DKZ22, we therefore need to rotate it to this
configuration.

There is, however, a small numerical subtlety. The first rotation, putting $\vec{l}$ into the positive $z$-direction, is a straightforward rotation around the
$\vec{l}\times \vec{e}_z$ axis, where $\vec{e}_z$ is the unit vector in the
global $z$-direction. In our 3D model we perform this rotation as a coordinate
mapping using linear interpolation, and, of course, we also rotate the velocity
vectors accordingly. The second rotation, however, which is the rotation around
the $z$-axis such that $\vec{\psi}$ points into positive $x$-direction, can cause
numerical difficulties. This is because numerical noise can make the direction
of the warp vector $\vec{\psi}$ wiggle wildly at radii for which $|\vec{\psi}|\ll
1$, e.g., around radii where it flips sign. As long as we only study a single
annulus of the disk this is not a problem. But when comparing the results of
neighbouring annuli, we might find major differences simply due to large
differences in horizontal ($x$-$y$) orientation. This is purely an artefact of the
choice of horizontal orientation. Since our disk model is initiated with a warp
purely in the $x$-$z$-plane, we choose, for this
section, to omit the second rotation. The global orientation of the model is
close enough to the DKZ22 choice that we can perform our analysis.

\begin{figure*}[ht!]
  \begin{center}
  \includegraphics[width=0.9\textwidth]{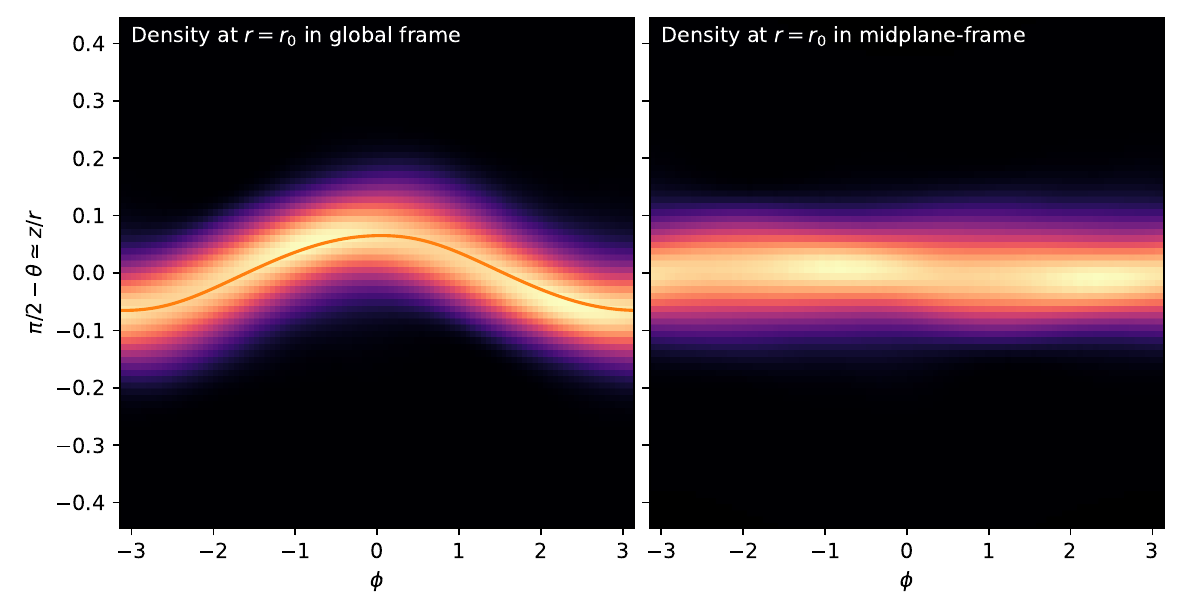}
  \end{center}
  \caption{\label{fig-density-annulus}Gas density of the disk annulus at $r=8.3\,\mathrm{au}$
    at time $t=120\,\mathrm{yr}$. The map is a standard $(\phi,\theta)$ projection
    of the sphere of constant distance to the star. The color scale is
    linear. Left: in unrotated coordinates.  Right: in rotated coordinates such
    that the disk annulus lies in the equatorial plane.}
\end{figure*}

We analyse our model at time $t = 120\,\mathrm{yr}$ (10 orbits at $R_0$),
at a radius of $r=8.3\,\mathrm{au}$.
We choose this point in time because the disk already had some time to relax and the internal torques are fully developed, but the disk system did not have time yet to change much.
At this moment, this disk annulus is
inclined with respect to the coordinate system by $-3.8\degree$. If we
display the density in the unrotated angular coordinates $\theta,\phi$, this
leads to the image Figure~\ref{fig-density-annulus}-left, where the annulus looks wavy. If we remap
this into a new, rotated, coordinate system, the disk annulus becomes
straight (Figure~\ref{fig-density-annulus}-right). One can see that the
density at the disk midplane has moderate maxima and minima. They are
caused by the vertical expansion and compression motion, called
``breathing motion'' by \citet{OgilvieLatter2013a}.

\begin{figure*}[ht!]
  \begin{center}
  \includegraphics[width=\textwidth]{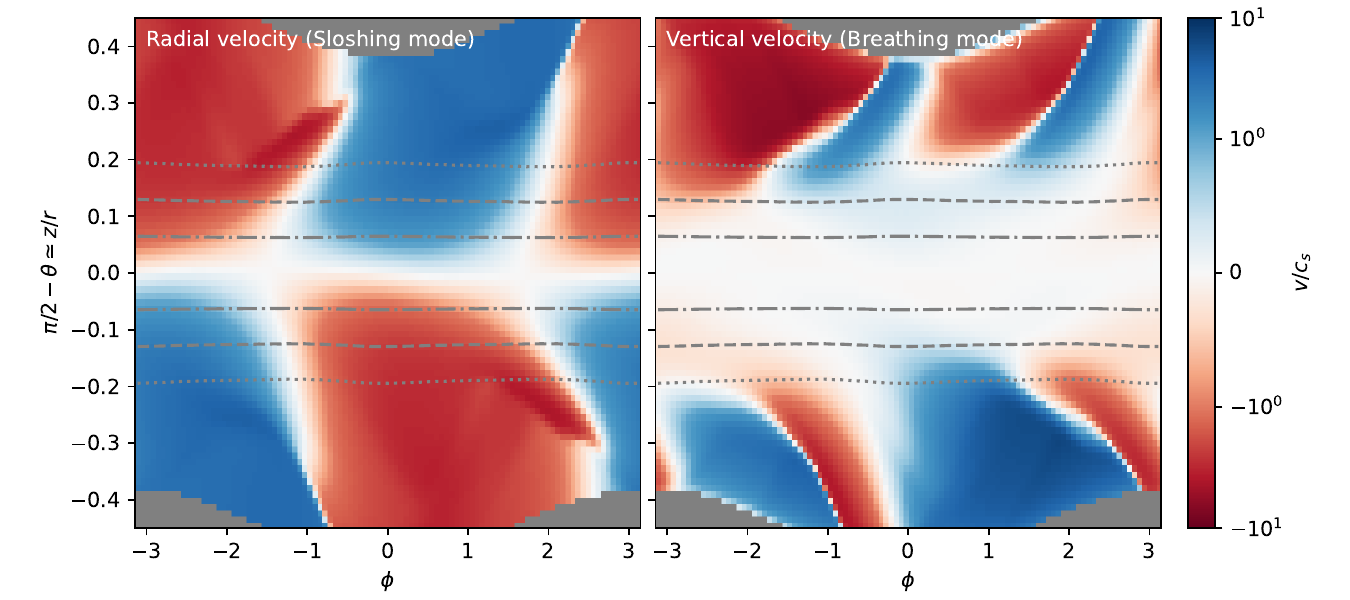}
  \end{center}
  \caption{\label{fig-velocities-annulus}Gas velocity components in units of the
    local isothermal sound speed at $r=8.3\,\mathrm{au}$ at time $t=120\,\mathrm{yr}$, in the
    rotated frame (as in Figure~\ref{fig-density-annulus}-right). The
    colors are linear between -1 and 1, and logarithmic for $|v/c_s|>1$. Note that
    the isothermal sound speed $c_s$ is constant at $0.66\,\mathrm{km/s}$ throughout each panel,
    consistent with a disk aspect ratio of $h_p/r=0.064$. Left:
    radial velocity (sloshing motion). Right: vertical velocity (breathing
    motion). The dot-dashed, dashed and dotted grey lines mark the \rev{root mean square vertical extend of the disk (which would relate to the pressure scale height in a steady-state disk)}, twice that, and three times that. \rev{The grey} areas below and
    above are the regions outside the model grid.}
\end{figure*}

In this new rotated frame, we now analyse the radial velocity
$v_r(\theta,\phi)$, i.e.\ the sloshing motion, and the vertical velocity
$v_z(\theta,\phi)$, i.e.\ the breathing motion. They are shown in
Figure~\ref{fig-velocities-annulus}. The Keplerian motion of the gas flows from left
to right (toward increasing $\phi$), which can therefore be regarded
as kind of a time axis, even though these figures show one snapshot in time.
In the left panel it can be seen that, as
expected, the radial velocity $v_r(z,\phi)$ has consistently opposite sign
above/below the disk midplane. This is the sloshing motion that produces
the internal torque. Far from the midplane, this sloshing motion becomes
supersonic. However, most of the torque is produced below about one
scale height \rev{(root mean square height of the disk above the disk's midplane)}, i.e.\ the grey dot-dashed line in the figure, where the
motions are subsonic, so that the
torque is presumably not much affected by the supersonic motions higher
up. \rev{The grey lines in Figure~\ref{fig-velocities-annulus} mark the vertical extend of the disk, i.e., the root mean square widths of the vertical Gaussian density profile. The lines become slightly wavy due to the breathing motion causing the vertical compression and expansion in the disk.}

The right panel shows the vertical motion of the gas. In this figure, the sign flips at
the disk midplane as well, meaning that the motion is either expanding or compressing
the disk vertically (i.e.\ the breathing motions). These motions are much weaker
than the sideways sloshing motions, but they also become supersonic far
above the disk midplane, leading to the formation of shocks where the upward
(blue) motion hits the downward (red) motion.

These results show that the sloshing and breathing motions predicted from
the linearised shearing box model \citep[][DKZ22]{OgilvieLatter2013a}
are indeed occurring in the 3D warped disk model. And these velocities can
be quite large, close to or exceeding the sound speed, especially at two
or more scale heights above the disk surface. This may have relevant
consequences for kinematic observations (using molecular lines) of
warped protoplanetary disks.

\begin{figure} [ht!]
  \includegraphics[width=0.5\textwidth]{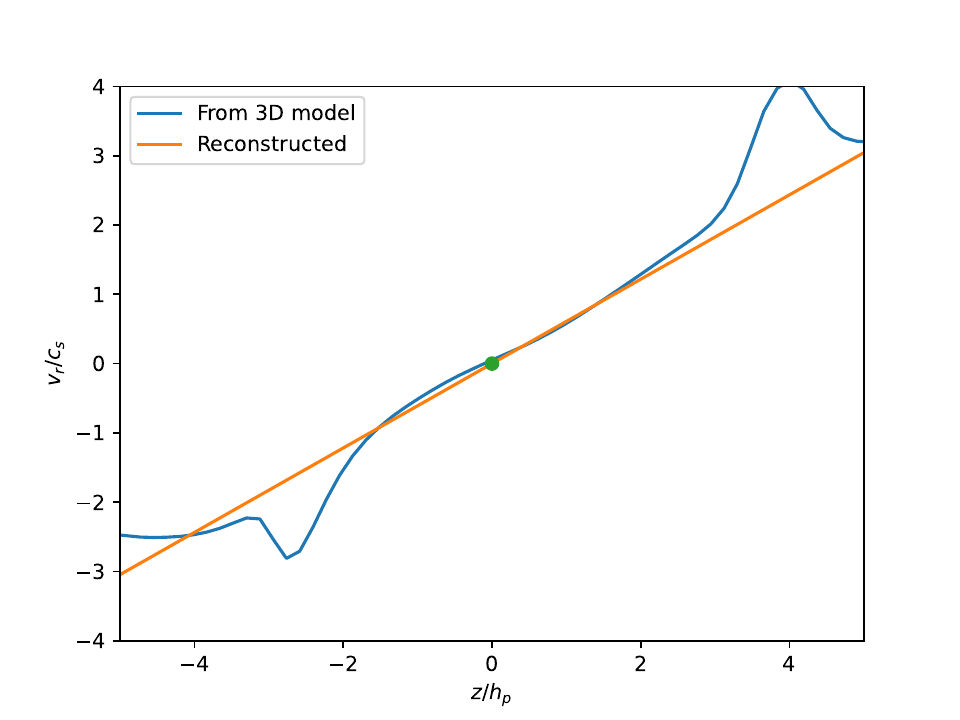}
  \caption{\label{fig-velocities-position}Radial gas velocity in units of the
    local isothermal sound speed, at the same radius and time as in
    Figure~\ref{fig-velocities-annulus}, at azimuth $\phi=1.29$, i.e., a vertical
    slice of that figure. The blue curve shows the data from the 3D model, the orange curve a
    linear reconstruction using a slope computed as in Appendix~\ref{sec-extracting-V}.}
\end{figure}

In Figure~\ref{fig-velocities-position} we show a vertical slice of
Figure~\ref{fig-velocities-annulus}-left, at azimuth $\phi=1.29$. The figure shows that
the linear approximation of the sloshing motion $v_r(r,z,t)=V_r(r,t)\Omega z$,
which stands at the basis of all 1D warped disk models, appears to be a good
approximation up to several scale heights at this radius and time. Deviations do
occur at some other radii and times, but overall they are minor.

\section{Conclusion} \label{sec:conclusion}

We performed three-dimensional simulations of misaligned disks using FARGO3D \citep{FARGO3D}.
To study the applicability of the grid-based code to misaligned features, we set up a disk \rev{that is} planar, but tilted with respect to the grid midplane and compared the result to a planar disk \rev{aligned to} the grid midplane.
The latter case has been studied extensively in previous works, therefore we use this as reference case.
Physically, the tilted case should lead to the same results, meaning all deviations in the simulation result from grid effects, i.e., numerical friction.
We find that grid-based methods can simulate misaligned features surprisingly well, given a high enough resolution.
We find the vertical resolution (in $\theta$) to be most important in avoiding grid effects.
Additionally, the minimum required resolution to accurately model misaligned features depends on disk viscosity: for higher viscosity, a higher resolution is needed.
We recommend for future studies using high viscosities in grid-based methods, to perform resolution studies in these specific cases.

We set up an initially warped disk around a single star and do not include any component driving the warp.
Evaluating the warp evolution in this isolated system, we find the evolution to be in a wave-like manner as expected.
Comparing the evolution to a one-dimensional ring code model, we find good agreement in inclination evolution.
Looking at the twist of the disk, however, we find a significant twisting in the three-dimensional model, which is not seen in the one-dimensional model.
The absence of this effect in the one-dimensional model does not necessarily mean that it is an unphysical effect.
Looking at the total angular momentum of the whole disk, it does not precess and is conserved in both absolute value and direction, which indicates the twist not to be caused by grid effects.
Further tests, including a comparison to one-dimensional models and decreasing the importance of the outer boundary by moving it away from the disk edge, point to the same conclusion.
For higher viscosity, we observe a less pronounced twisting with a faster damping of the precession motion.
We suspect the twist to be a physical effect in the three-dimensional warp evolution.
The reason why it is not seen in one-dimensional models could be that the effect was neglected in the linearisation process.

Evaluating the internal dynamics in our simulation in detail, we find the sloshing and breathing motion as predicted in the local shearing box approach in DKZ22.
These motions are mainly responsible for the warp evolution and generate the internal torque.
We find that the sloshing and breathing velocities can become supersonic in the upper layers of the disk.
Although the upper layers do not greatly influence the warp evolution, as the internal torque is produced close to the midplane, it can have a significant influence on kinematic observations of warped disks.

\begin{acknowledgements}
We kindly thank Philipp Weber and Thomas Rometsch for their help in using FARGO3D and the bwForCluster BinAC, inspiration of plots, and comments on the project.
\rev{We thank Rebecca Martin for helpful discussions.}
We further thank Hubert Klahr and Ralf Klessen for discussions and helpful comments.
We remember the legacy of Prof. Willy Kley, who passed away in 2021, and would have been a part of this work.
\rev{Additionally, we thank the referee for very helpful comments and suggestions.}
We acknowledge funding from the DFG research group FOR 2634 “Planet Formation Witnesses and Probes: Transition Disks” under grant DU 414/23-2 and KL 650/29-1, 650/29-2, 650/30-1.
Additionally, we acknowledge support by the High Performance and Cloud Computing Group at the Zentrum für Datenverarbeitung of the University of Tübingen, the state of Baden-Württemberg through bwHPC and the German Research Foundation (DFG) through grant INST 37/935-1 FUGG.
We gained great benefit for this work from Core2disk-III, which is supported by the LABEX P2IO, the PCMI, PNPS, and PNP programs, and the OSU Paris-Saclay.
Plots in this paper were made with the Python \citep{Python3} library \texttt{matplotlib} \citep{matplotlib}. We acknowledge also the usage of \texttt{numpy} \citep{NumPy}, \texttt{scipy} \citep{SciPy}, and \texttt{astropy} \citep{AstroPy}.

\end{acknowledgements}

\bibliographystyle{aa}
\bibliography{3d-warp} 

\begin{appendix}

\section{\rev{A High Resolution Simulation of a Warped Disk}}\label{sec:appendix-highres}

\rev{In order to check the dependency on resolution, we performed a simulation of our fiducial warped disk set-up with a higher resolution.
For this simulation, we doubled the resolution in all three directions, which means 160 cells in $r$-direction (3.4~cells-per-scaleheight at $r=5.2\,\mathrm{au}$), 200 cells in $\phi$-direction (1.6~cps at $r=5.2\,\mathrm{au}$) and 264 cells\footnote{\rev{The resolution in $\theta$ should actually have been 268 cells for a doubled resolution, but due to a miscalculation, we set 264. However, we do not expect any significant changes in the result and therefore decided not to run the simulation again.}} (8.2~cps at $r=5.2\,\mathrm{au}$) in $\theta$-direction.}

\begin{figure} [ht!]
  \centerline{\includegraphics[width=0.5\textwidth]{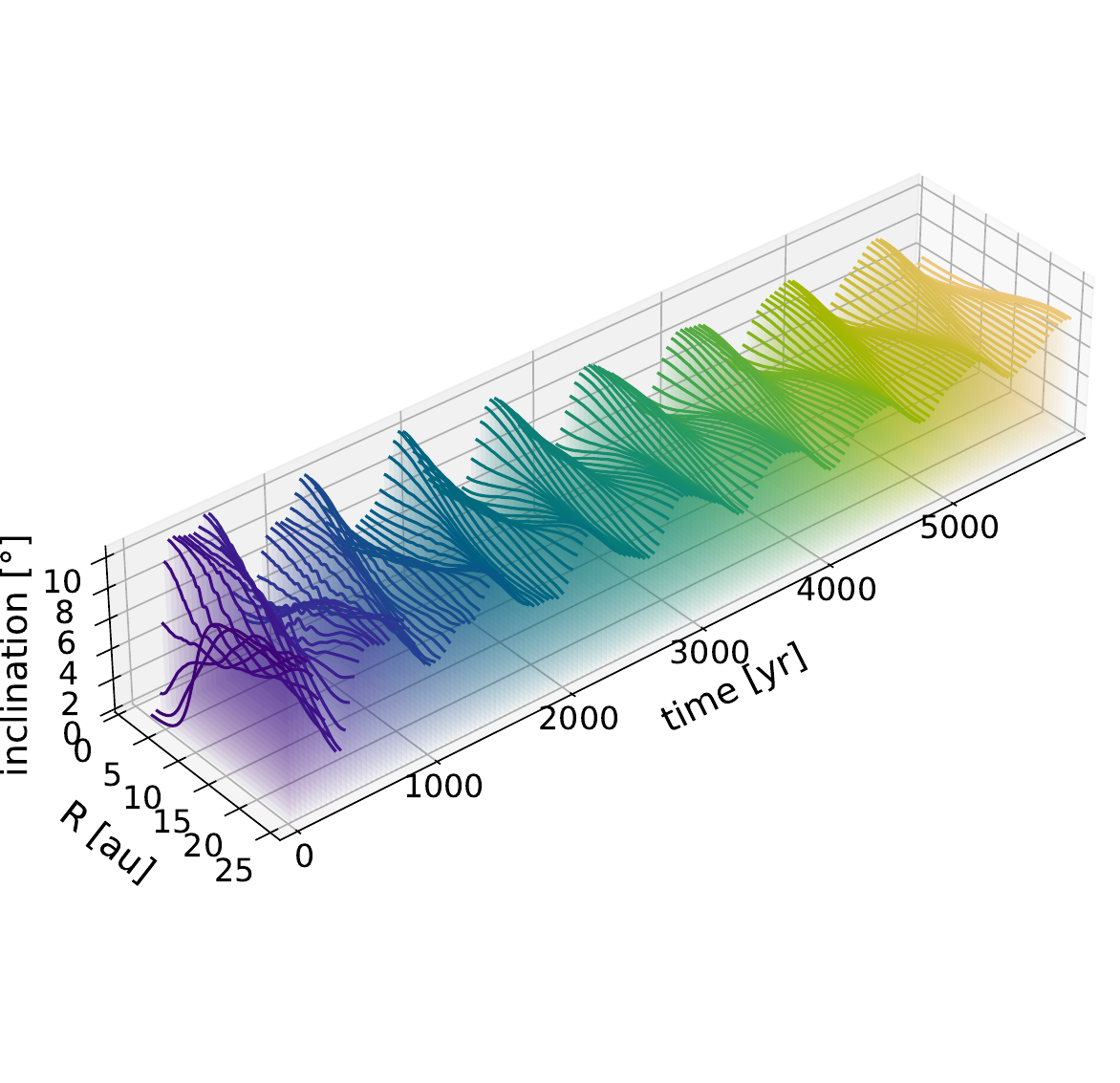}}
  \caption{\label{fig:warp-incl-highres} \rev{Evolution of the inclination profile in the high-resolution warped disk simulation. The color highlights the time. Similar to Figure~\ref{fig:warp-incl}.}}
\end{figure}

\rev{Figure~\ref{fig:warp-incl-highres} shows the wave-like behavior of the simulation. Like in the simulation with lower resolution, we see a global bending mode superimposed on a global tilt. The bending mode appears as standing wave with a wavelength of twice the radial disk regime.
}

\begin{figure} [ht!]
  \centerline{\includegraphics[width=0.5\textwidth]{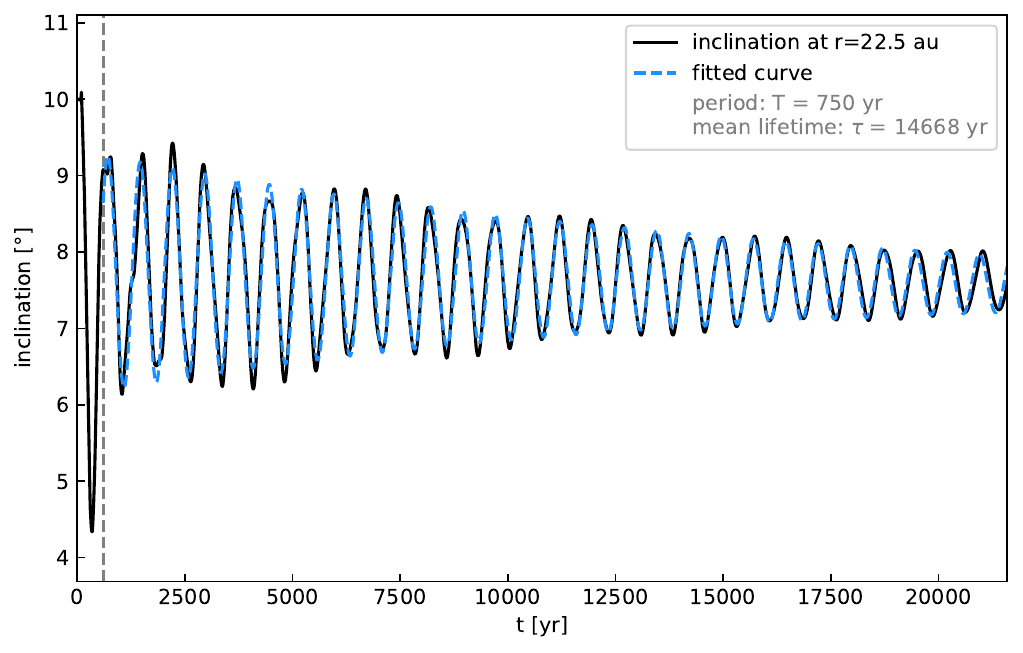}}
  \caption{\label{fig:incl-decay-highres} \rev{Decay of inclination in the high resolution warped disk simulation. We evaluate the inclination at $r = 22.5\,\mathrm{au}$. The black line shows the simulation data, the blue dashed line shows the fit. The grey dashed line indicates the starting point of the fit. Similar to Figure~\ref{fig:incl-decay} (note the slightly shorter simulation time).}}
\end{figure}

\rev{To evaluate the bending mode in detail, we fit the inclination evolution in the outer part of the disk (i.e., at $r=22.5\,\mathrm{au}$) with Equation~\ref{eq:fit} (see Figure~\ref{fig:incl-decay-highres}). The fit parameters give an amplitude $A = 1.6\degree$, a frequency of $\omega = 0.0083\,\mathrm{yr}^{-1}$, a decay rate of $\lambda = 6.8 \cdot 10^{-5}\,\mathrm{yr}^{-1}$, a global tilt of $c = 7.7\degree$, and a global (numerical) inclination loss of $d = -6.8 \cdot 10^{-6}\,\mathrm{yr}^{-1}$.
These parameters indicate a period of $P = 750\,\mathrm{yr}$ and a mean lifetime of $\tau = 1.5\cdot 10^4\,\mathrm{yr}$.
}

\rev{In comparison to the fiducial resolution simulation, we find a slightly shorter wave period and a slightly lower decay rate leading to a slightly longer mean lifetime of the warp. However, the difference is small (much lower than one order of magnitude) and we can therefore say that the two simulations compare well with each other.
Unsurprisingly, we find a much lower inclination loss $d$ by one order of magnitude, which supports our assumption that this is caused by numerical effects.
}

\begin{figure} [ht!]
  \centerline{\includegraphics[width=0.5\textwidth]{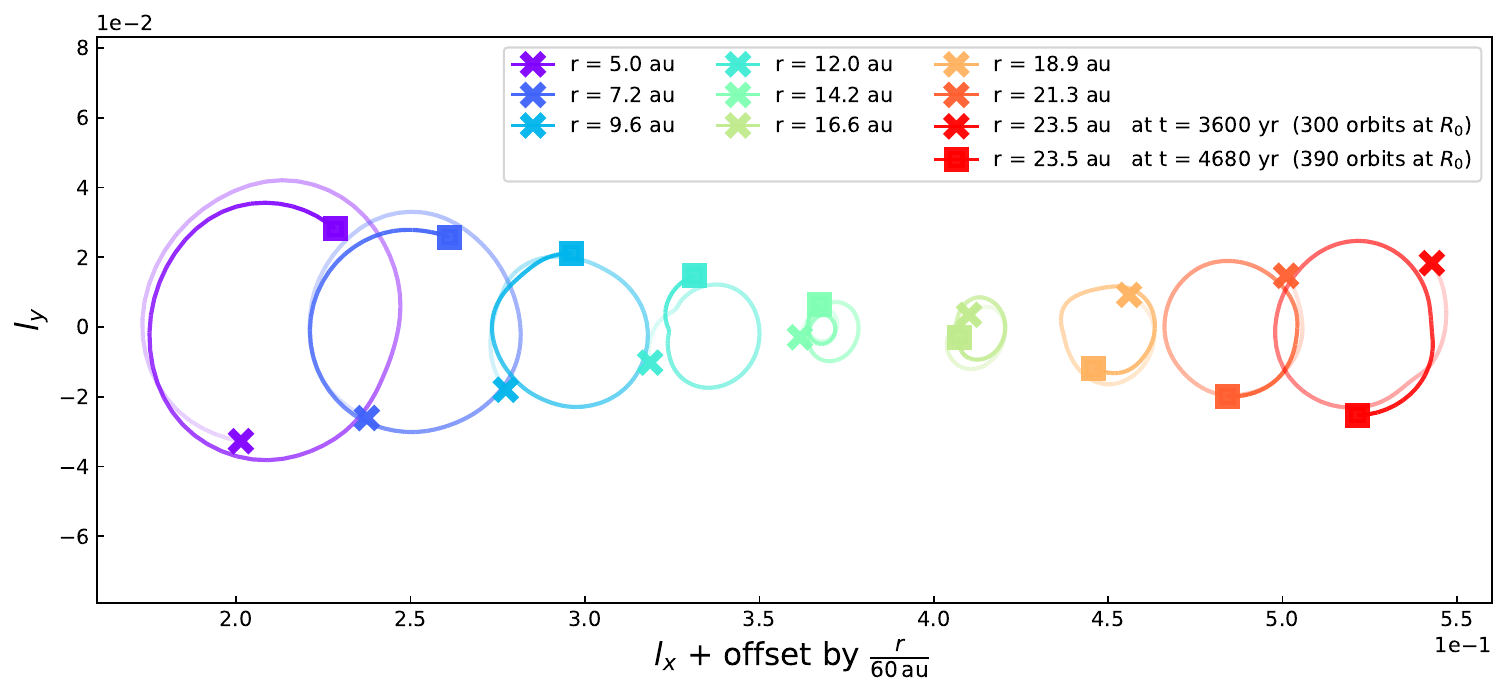}}
  \caption{\label{fig:twist-highres} \rev{Twist of the disk in the high resolution simulation (like Figure~\ref{fig:precession-radii}, bottom panel).}
  }
\end{figure}

\rev{Additionally, we briefly evaluate the twisting of the disk in the high resolution simulation, see Figure~\ref{fig:twist-highres}.
For the plot, we choose the same time intervals as for the fiducial resolution.
The twisting behavior of the disk stays the same and does not seem to be strongly influenced by resolution.
}

\begin{figure} [ht!]
  \centerline{\includegraphics[width=0.5\textwidth]{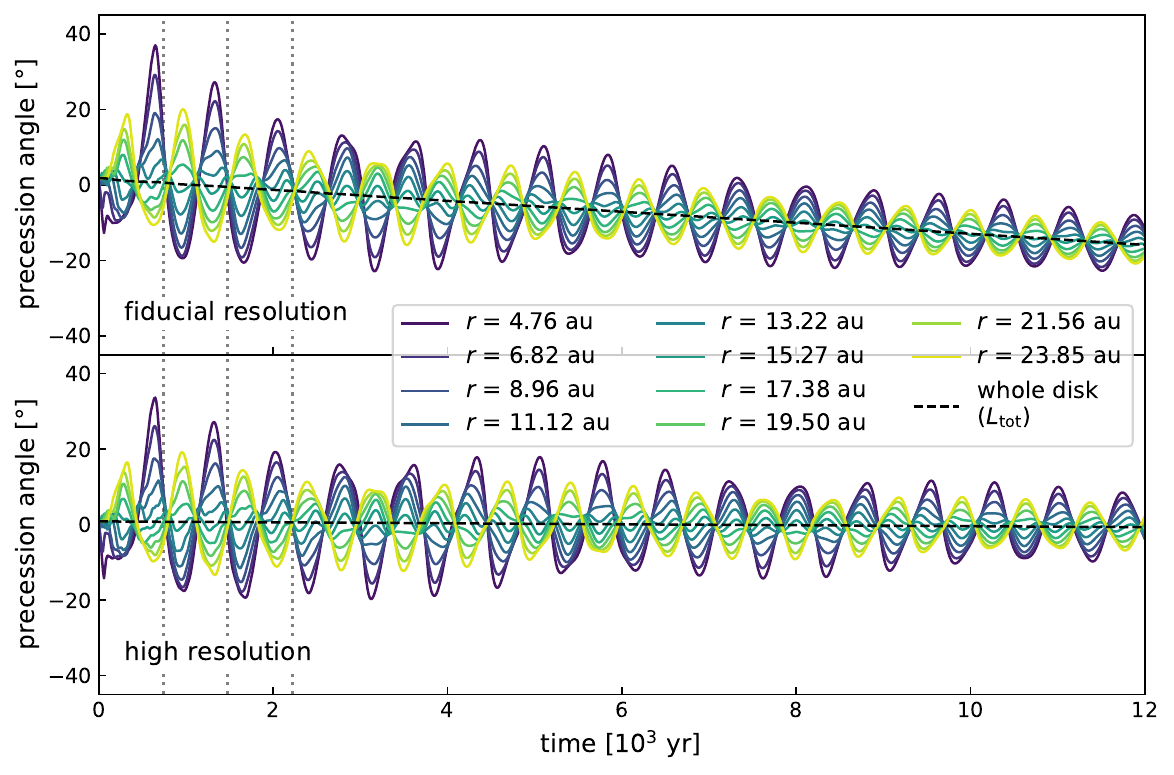}}
  \caption{\label{fig:prec-highres} \rev{Precession angles for different radii (colored lines) in simulations with different resolutions. The black dashed line in each panel shows the precession angle of the angular momentum vector $\vec{L}_\mathrm{tot}$ of the whole disk. The grey dotted lines indicate precession periods of roughly $740\,\mathrm{yr}$. Similar to Figure~\ref{fig:prec-viscosities}.}
  }
\end{figure}

\rev{Figure~\ref{fig:prec-highres} shows the evolution of the precession angles at different radii, as well as the overall precession of the disk in both the fiducial and high resolution simulations.
The comparison clearly shows that the overall precession of the whole system (negative slope of the black dashed line for the fiducial resolution) is an effect caused by numerical effects, as it does not decrease significantly in the high resolution simulation.
The twisting motion, however, is very similar in both simulations, suggesting that it is not caused by numerical effects due to too low resolution.
The estimated period of the twisting motion of roughly $740\,\mathrm{yr}$ (grey dotted lines) coincides with the fitted period of inclination evolution.
}

\section{\rev{Investigation of the Disk Twist}}\label{sec:appendix-twist}

\rev{In this section, we perform a few tests in order to investigate whether the twist we find in the warped disk simulation is a numerical or a physical effect.}

\subsection{Outer Boundary}

We \rev{first} perform another three-dimensional simulation \rev{where the outer boundary is far away from the disk edge by} applying an exponential cut-off to the density profile.
\rev{In this case,} the grid edge \rev{is unlikely to influence the simulation.}

\begin{figure} [ht!]
  \centerline{\includegraphics[width=0.5\textwidth]{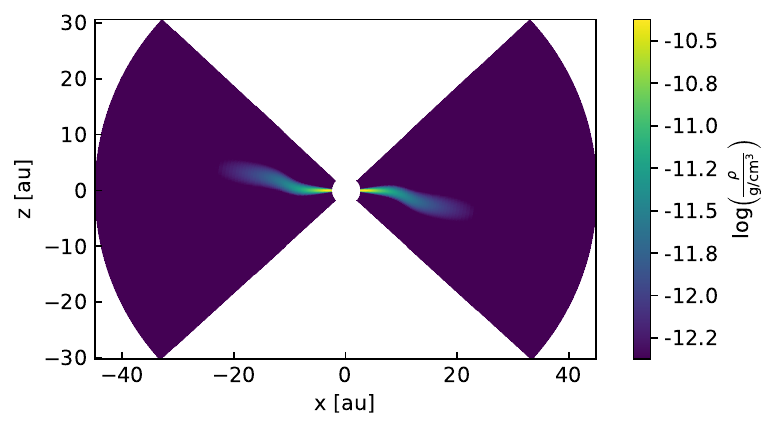}}
  \caption{\label{fig:setup-outbound} \rev{Set-up} of a warped disk with the outer boundary further away from the physical disk edge.
  }
\end{figure}

Figure \ref{fig:setup-outbound} shows the initial \rev{set-up} of this simulation.
We keep the same initial conditions as in the previous simulation of the warped disk, the only difference being the cut-off at $r=26\,\mathrm{au}$ in the density profile.
As we enlarge the radial space of the grid, we keep the number of cells in the disk domain (within 26\,au) the same.

\begin{figure} [ht!]
  \centerline{\includegraphics[width=0.5\textwidth]{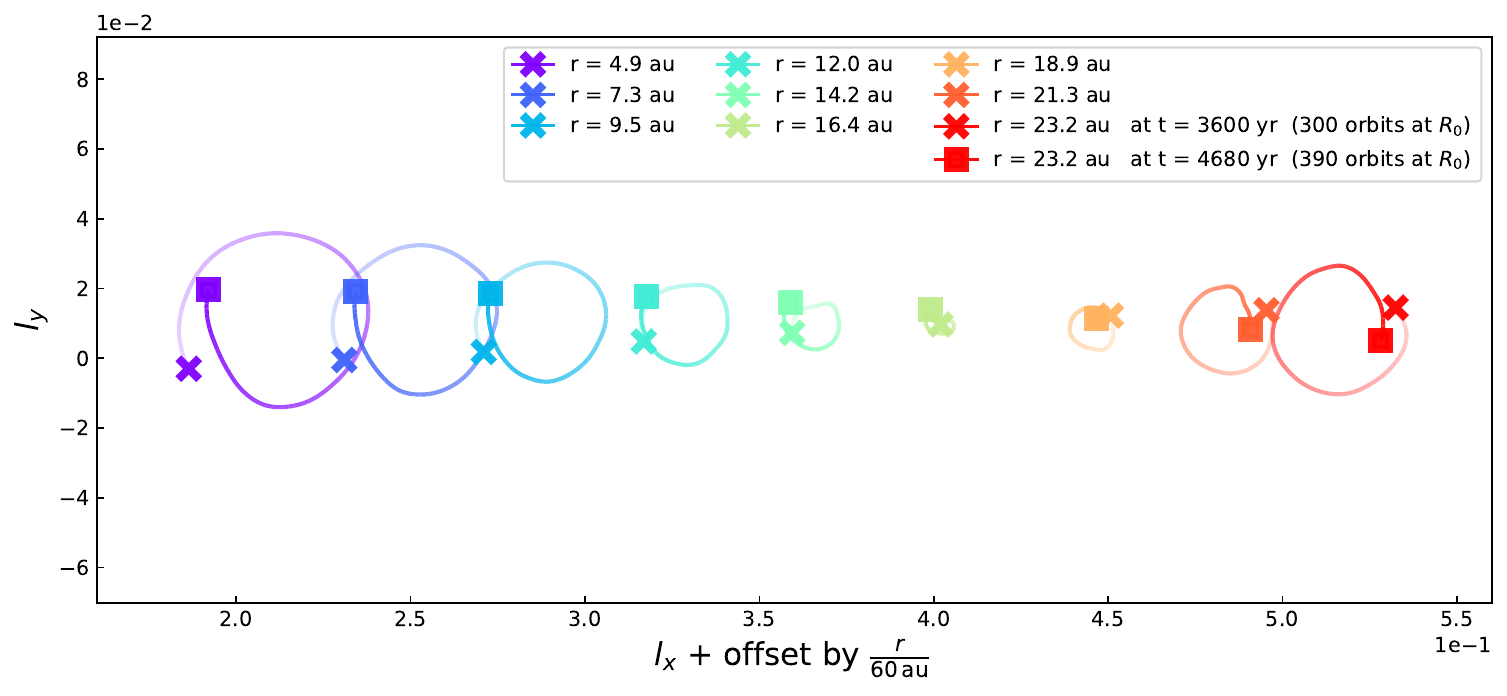}}
  \caption{\label{fig:twist-outbound} Twist of a disk with the outer boundary far away from the disk edge, like Figure \ref{fig:precession-radii}, bottom panel.
  }
\end{figure}

\begin{figure} [ht!]
  \centerline{\includegraphics[width=0.5\textwidth]{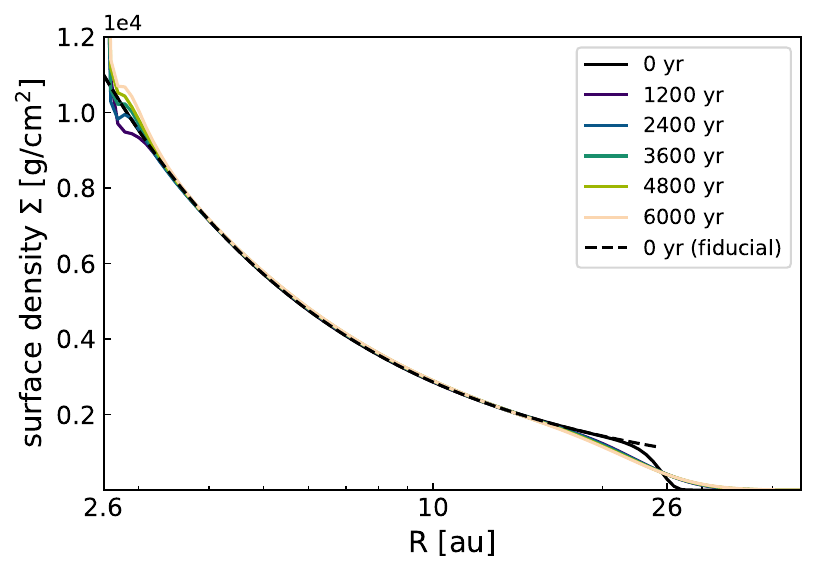}}
  \caption{\label{fig:surfdens-outbound} Surface density of a disk with the outer boundary far away from the disk edge (solid lines). The initial state of the fiducial simulation is plotted as reference (dashed line).
  }
\end{figure}

Figure~\ref{fig:twist-outbound} shows that we \rev{also} find twisting of the disk.
However, the frequency is on a longer time scale than before.
Evaluating the inclination evolution of the disk in this \rev{set-up}, we find that it also takes place on longer time scales, i.e. that the warp wave takes longer to travel through the disk.
Again, the time scale of the inclination wave corresponds with the time scale of the twist frequency.
We find the reason for this to be the cut-off smoothing out on relatively short time scales, as shown in Figure~\ref{fig:surfdens-outbound}.
This leads to the disk being slightly larger, therefore the time scale of the warp wave is larger.
\rev{We suspect the reason for this to lie in our initial set-up of the azimuthal velocity, as the smooth-out occurs very quickly. In future work, this could be improved by accounting for the density gradients in the initial set-up.}

However, the outer boundary is still unlikely to influence the outer disk edge in this simulation, because the density is very low outside of $26\,\mathrm{au}$.
The fact that the twisting persists is an indication that the twist results from a physical process in a warped disk.

\subsection{\rev{1D Models}}

In order to rule out that the twist is an effect kicked off some time in the beginning of the three-dimensional simulation, we investigate the behaviour of a twist in a one-dimensional model.
For this, we take the twisted state at 308 orbits at $R_0$, which is $t~=~3696\,\mathrm{yr}$ and use it as initial condition for a one-dimensional simulation using our code \texttt{dwarpy} (Kimmig et al., in prep).

\begin{figure} [ht!]
  \centerline{\includegraphics[width=0.5\textwidth]{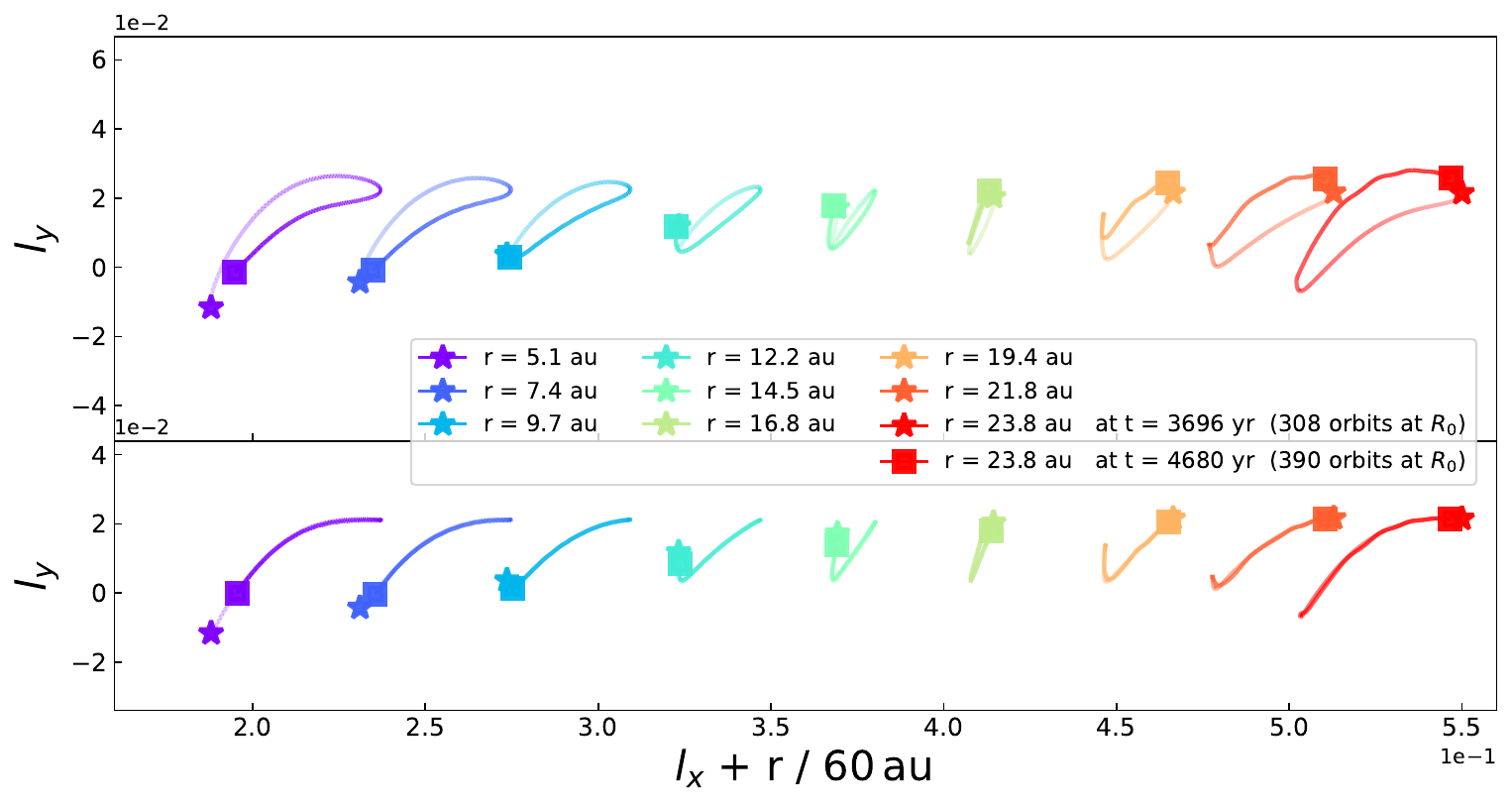}}
  \caption{\label{fig:twist-dwarpy} Evolution of the twisted state in a one-dimensional simulation using \texttt{dwarpy} in two different simulations. The initial state of the warp in both simulations is taken from the twisted state in the three-dimensional simulation at $t~=~3696\,\mathrm{yr}$ (308 orbits at $R_0$).
  The top panel shows the 1D simulation with internal torque $\vec{G}$ extracted from the 3D simulation, the bottom panel shows the same simulation, but internal torque $\vec{G}$ initially set to zero for comparison.
  Comparable to Figure \ref{fig:precession-radii}, bottom panel.
  }
\end{figure}

We perform two simulations using this state of the warp. In the first simulation, we use the internal torque extracted from the three-dimensional simulation (see Appendix~\ref{sec-internal-torques}).
To evaluate the difference it makes, we perform another simulation setting the internal torque values initially to zero and let the one-dimensional equations take care of its evolution.
The first case (with extracted internal torque) is shown in Figure \ref{fig:twist-dwarpy}, top panel, the latter case (internal torque set to zero) in the bottom panel.
In both panels, the stars show the initial state of the unit angular momentum vectors (same as the grey stars in the bottom panel of Figure~\ref{fig:precession-radii}).
The squares show the state at the end of our simulation time span, which we chose to be the same as plotted in Figure~\ref{fig:precession-radii}.

We find that the one-dimensional model fails to recreate the circular motion of the unit angular momentum vectors.
The simulation with the initial torque set to zero (bottom panel of Figure~\ref{fig:precession-radii}) shows no circular motion at all, while the simulation with internal torque extracted from the three-dimensional simulations shows deformed ellipses.
The latter is the more self-consistent one, which shows a hint of the circular twist motion, but can not reproduce the full twist seen in the three-dimensional simulation.

From this, we can say that the twisting of the disk seems to be a three-dimensional effect which is not captured in the one-dimensional equations.
It might be an effect traced by \rev{linear or} non-linear terms that were neglected in deriving the equations.
We can conclude here that the twist is an effect triggered continuously in the three-dimensional simulation, as opposed to an event kicked-off once.

\subsection{\rev{Keplerianity} of the Warped Disk}

In one-dimensional models, an untwisted and perfectly Keplerian disk will not twist.
However, if the disk deviates from \rev{Keplerianity}, a twist can be triggered (second term on left hand side of Equation~\ref{eq:internal-torque}).
We therefore take a closer look at the azimuthal velocity of the warped disk in the three-dimensional model.
Because the orbital plane does not always align with the grid plane, the evaluation of the azimuthal velocity is not straight forward: we need to perform an interpolation.
For the interpolation, we use the unit angular momentum vector determining the orbital plane of each grid shell and use the \texttt{LinearNDInterpolator} function from \texttt{scipy} \citep{SciPy}.
This way, we can extract the azimuthal velocity of the disk midplane in each grid shell.

\begin{figure} [ht!]
  \centerline{\includegraphics[width=0.5\textwidth]{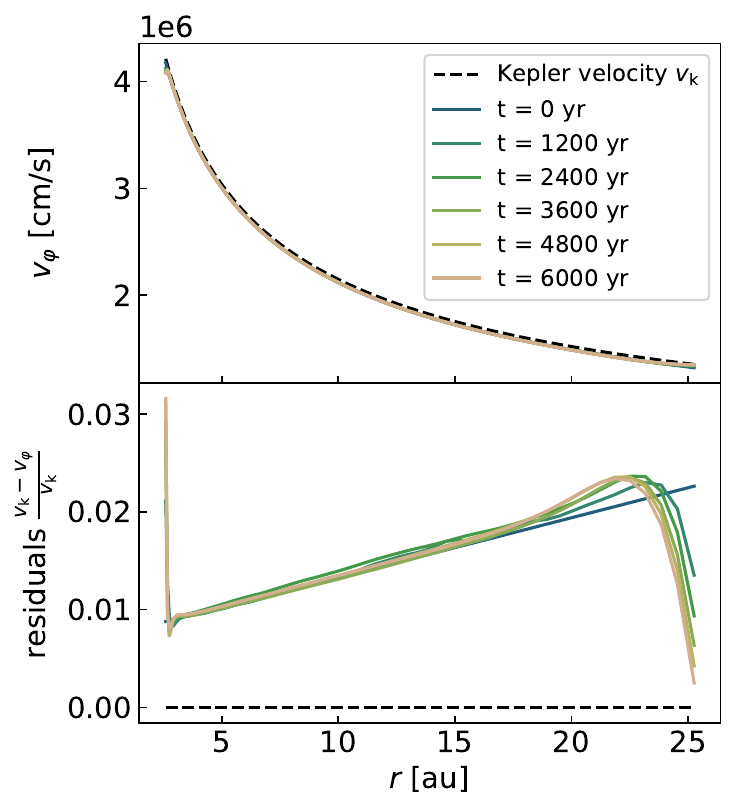}}
  \caption{\label{fig:keplerity} Azimuthal velocity (top panel) and the residuals from the Kepler velocity (bottom panel). The black dashed line shows the Kepler velocity.
  }
\end{figure}

Figure \ref{fig:keplerity} shows that the azimuthal velocity of the disk deviates from the Kepler velocity as we set up the disk with pressure correction terms.
During the simulation, we do not observe strong deviations from our initial \rev{set-up}, except at the inner and outer boundaries, which means that the warp does not significantly change the azimuthal velocity.
The deviation from Keplerian motion results from pressure correction in the initial \rev{set-up} of the disk.
Because the deviation is small, we do not expect this to be the sole reason for the twist.
In future work, this could be investigated in one-dimensional models with an adjusted apsidal frequency.


\subsection{\rev{Twist of a Differently Flared Disk}}
\rev{In this section, we investigate the twist behavior of a locally isothermal disk with a flaring index of $i_\mathrm{fl} = 0.25$, as described in Section~\ref{sec:fl25}.}

\begin{figure} [ht!]
  \centerline{\includegraphics[width=0.5\textwidth]{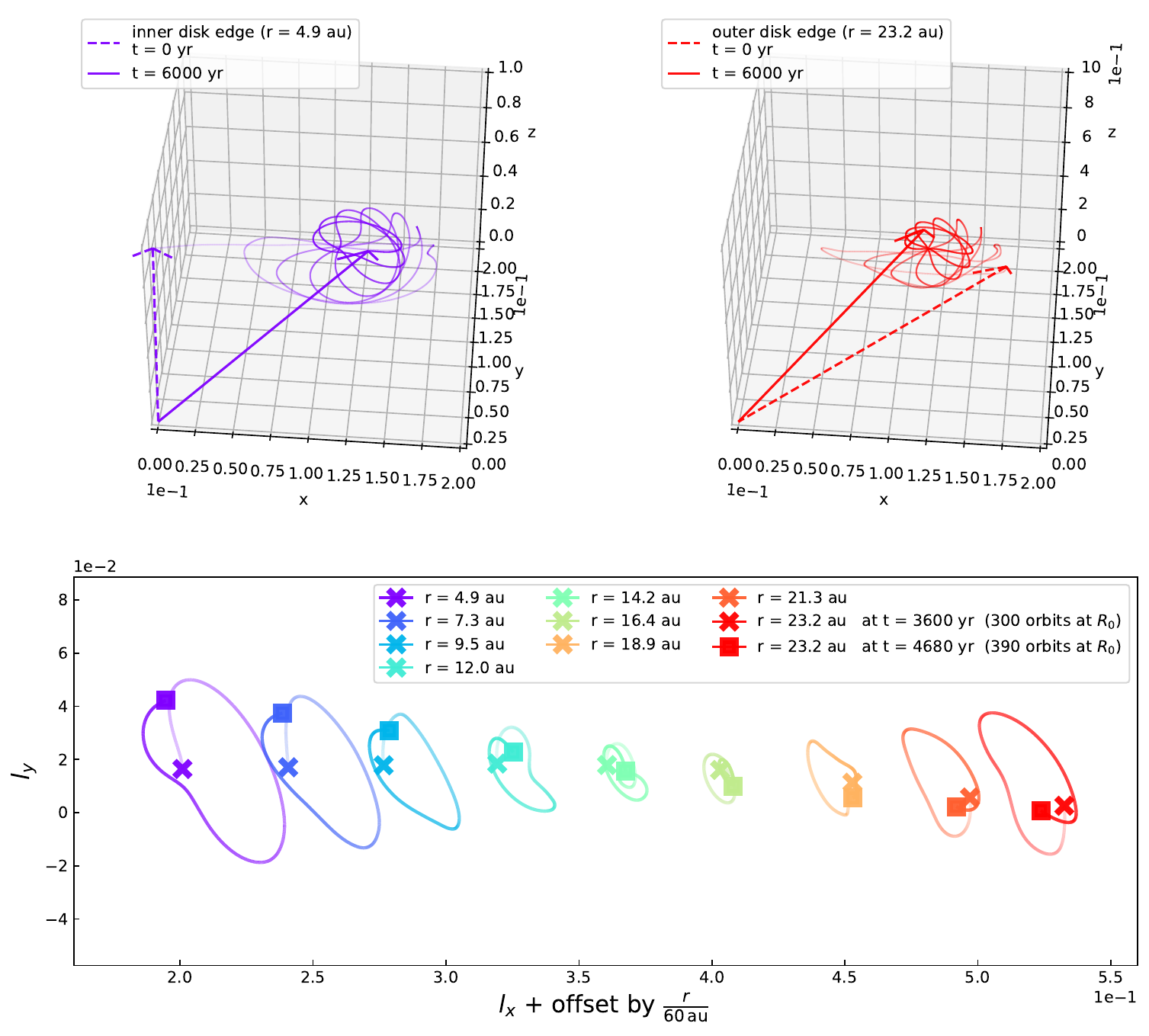}}
  \caption{\label{fig:precession-radii-fl25} \rev{Twist motion in the locally isothermal disk. \textbf{Top:} Unit angular momentum vectors close to the inner (left panel) and outer (right panel) edge of the disk for the first $t~=~6000\,\mathrm{yr}$.
  \textbf{Bottom:} $x$-$y$-components of the unit angular momentum vector for the time period between 300-390 orbits at $R_0$ at varying radii, plotted with an offset of $r/(60\,\mathrm{au})$ in $y$-direction to disentangle the twist at different radii. The x-markers show the beginning of the plotted time, the squares the end.
  Similar to Figure~\ref{fig:precession-radii}.
  }}
\end{figure}

\rev{Figure~\ref{fig:precession-radii-fl25} shows the evolution of the unit angular momentum vectors. The top row indicates the time evolution of the vectors. We clearly see that a twist develops in the disk. However, the behavior looks more complicated than in the globally isothermal case (see Figure~\ref{fig:precession-radii}).
We find that instead of circles, the precession motion takes on an oval or crescent shape, whose orientation also precesses.
In the bottom panel, we plot the same time period as plotted in Figure~\ref{fig:precession-radii} in order to see the oval shape of the twisting motion more clearly.
}

\rev{We additionally checked the evolution of the total angular momentum vector in the locally isothermal case. Just like in the globally isothermal case, we find very little loss of absolute value ($0.67 \%$ within $2.4\cdot 10^4\,\mathrm{yr}$) and a slight change in direction ($4.2\degree$ within $2.4\cdot 10^4\,\mathrm{yr}$), which we suspect to occur due to the non-conservative nature of the code for out-of-plane features.
Most importantly, we also do not see any precession behavior of the total angular momentum vector, which means that the twist in the locally isothermal case is also conserving the total angular momentum within the disk.
}

\rev{We take the fact that a twist also occurs in a differently flared disk as another indication that this effect might be physical.
In further work, the influence of the vertical disk structure on the warp and twist behavior could be investigated in more detail.
}

\section{Internal Torques in the 3D Model}\label{sec-internal-torques}
From the 3D simulation of the disk we can, at any given time $t$, compute the
internal torques. We have the variables $\rho(r,\theta,\phi)$ for density,
$u_r(r,\theta,\phi)$, $u_\theta(r,\theta,\phi)$, $u_\phi(r,\theta,\phi)$ for the velocities and the
isothermal sound speed $c_s(r)$ at our disposal. The velocity components
$u_\phi$ and $u_\theta$ contain the Kepler rotation as well. To make the task of
computing the internal torque easier, we re-express the velocity in terms of
cartesian coordinates $X$, $Y$ and $Z$:
\begin{equation}
\left(\begin{matrix}
u_X \\ u_Y \\ u_Z
\end{matrix}\right)
= 
\left(\begin{matrix}
  \sin(\theta)\cos(\phi) & \cos(\theta)\cos(\phi) & -\sin(\phi) \\
  \sin(\theta)\sin(\phi) & \cos(\theta)\sin(\phi) &  \cos(\phi) \\
  \cos(\theta)           & -\sin(\theta)          & 0
\end{matrix}\right)
\left(\begin{matrix}
u_r \\ u_\theta \\ u_\phi
\end{matrix}\right).
\end{equation}
From this point onward, we will keep the spherical coordinate system
$(r,\theta,\phi)$ and the corresponding spherical grid, but express all vector
and tensor components in a vector basis tangent to the global cartesian
coordinates $(X,Y,Z)$. At any given point $(r,\theta,\phi)$ in our model we can
then define the stress tensor
\begin{equation}\label{eq-stress-tensor}
T_{ij} = \rho u_i u_j + p\delta_{ij},
\end{equation}
where we use the pressure $p=\rho c_s^2$, the Kronecker-delta $\delta_{ij}$, and
where we ignore viscosity. Here the indices $i$ and $j$ denote the $X$, $Y$
or $Z$ direction. We can now compute the local internal torque tensor at every
point in our 3D model:
\begin{equation}
g_{ij} = \varepsilon_{ikl}r_kT_{lj}
\end{equation}
(cf.~Eq.~110 of DKZ22, but now in the global cartesian basis), where
$\varepsilon_{ikl}$ is the Levi-Civita tensor, and where we use Einstein's
summation convention. The symbol $r_k$ is the position vector pointing from
the origin (i.e.\ the star) to the location in the 3D model where we compute
$g_{ij}$. Next we compute the internal torque vector, which is the radial
component of this tensor:
\begin{equation}
g_{i} = g_{ij}r_j/|r|
\end{equation}
(cf.~Eq.~111-113 of DKZ22), where $|r|$ denotes the length of the position
vector $r_j$. In contrast to DKZ22, we here use the spherically-outward
component rather than the cylindrically-outward component, because in our 3D
model we consider the warped disk as a series of spherical shells that mutually
exert torques on each other. The differences are, however, marginal.

\begin{figure} [ht!]
  \includegraphics[width=0.5\textwidth]{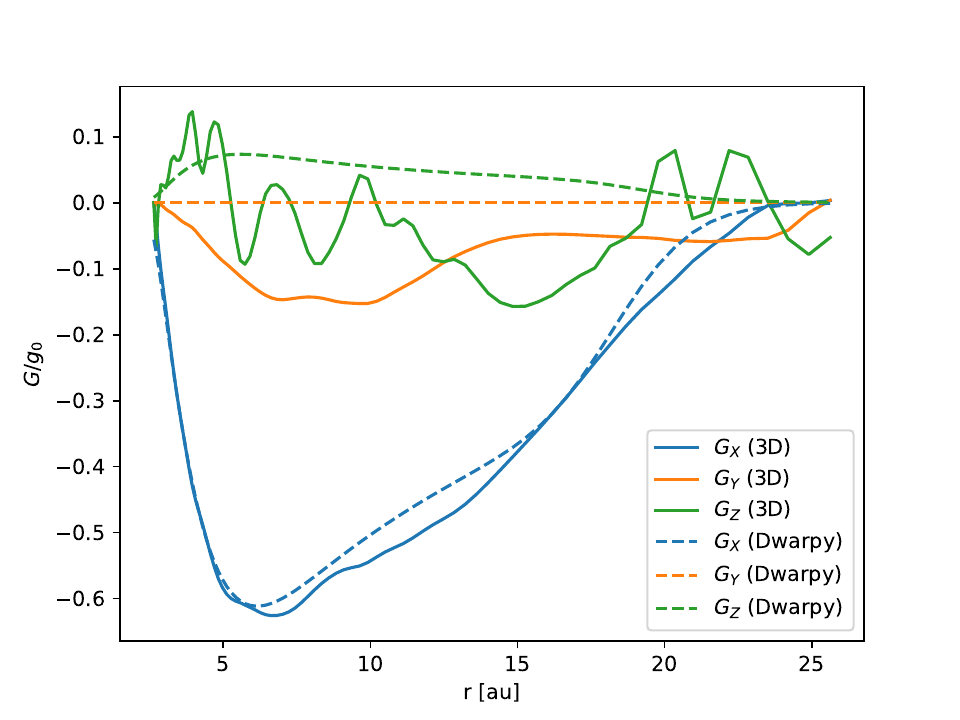}
  \caption{\label{fig-torques}The internal torques $G_x$, $G_Y$ and $G_Z$ at
    time $t=120\,\mathrm{yr}$ in units of $g_0=\Omega^2r\Sigma h_p^2$. Solid lines:
  from the 3D model. Dashed lines: from the 1D model \texttt{dwarpy}.}
\end{figure}

The local torque vector $g_i(r,\theta,\phi)$ is still a function of all
three coordinates. For the 1D warped disk equations, to which we want to
compare our 3D model, we need to compute, at each radius $r$, the integral
of this quantity over the sphere with radius $r$:
\begin{equation}\label{eq-Gi-torque}
G_i(r) = \frac{r}{2\pi}\int_{-\pi}^{\pi}d\phi\,\int_{-1}^{+1}d\cos(\theta) \; g_i(r,\theta,\phi).
\end{equation}
This is the internal torque vector that we can compare to the internal
torque from a 1D warped disk model (e.g., Eq.~173 of DKZ22). In Figure~\ref{fig-torques}
we show these torques in units of $g_0=\Omega^2r\Sigma h_p^2$ (see Eq.~126 of
DKZ22). The results show that the $G_X(r)$ of the 3D and 1D models match well.
This is the torque that is primarily responsible for the wavelike propagation
of the warp. The torque $G_Y$ is the torque that produces a twist. In the 1D
model this is perfectly 0, while in the 3D model it is small, but non-zero.
The $G_Z$ torque component in the 3D model is of the same order of magnitude
as $G_Y$, and is rather wiggly. Also in the 1D model it is not perfectly
zero, which is due to the fact that the disk is not perfectly aligned with
the equatorial $X-Y$ plane.

\section{Extracting the $V_r$, $V_\phi$ and  $V_z$ From the 3D Model}\label{sec-extracting-V}
As shown in DKZ22, and following \citet{OgilvieLatter2013a}, the sloshing
motion can be approximated as a linear velocity model
\begin{equation}
v_r = V_r \Omega z,\quad v_\phi = V_\phi \Omega z,\quad v_z = V_z \Omega z,
\end{equation}
where $v_z=-v_\theta$ by convention. In reality, the functions $v_r(z)$,
$v_\phi(z)$, and $v_z(z)$ are only approximately linear in $z$ close to the
midplane, but strongly deviate from linearity far away from the midplane, 
as shown for example in Figure~\ref{fig-velocities-position}.
However, usually the strongest deviation occur well
above one scale height, while the strongest effect on the disk dynamics is
where the density is the highest. A good estimate of $V$ for a velocity $v(z)$
is
\begin{equation}
V(r,\phi) = \frac{1}{\Sigma H_p^2\Omega}\int_{-\infty}^{+\infty} \rho(r,\phi,z) v(r,\phi,z) z dz.
\end{equation}
The $V_r(r,\phi)$ is the key component determining the torque. For each $r$ it
can be well approximated by
\begin{equation}
V_r(r,\phi) = V_{r0}(r) \cos(\phi-\phi_{r0}(r)),
\end{equation}
where $V_{r0}(r)$ is the amplitude of the sloshing motion, and
$\phi_{r0}(r)$ is the phase shift. As shown in DKZ22, for
a Keplerian disk in the wavelike regime, the phase shift, at least in the
linear theory, is expected to be $\phi_{r0}(r)=0$ if the warp vector is
pointing in the $\phi=0$ direction. In that case, the warp evolution is
entirely along that same direction, and the disk does not acquire a twist.
However, if $\phi_{r0}(r)\neq 0$, \rev{the disk} will acquire a twist, in
addition to the wavelike propagation of the warp itself.

From the 3D flattened model we can determine $V_{r0}(r)$ and $\phi_{r0}(r)$
in the following way: First consider, for each given radius $r$, 
\begin{equation}
V_r(\phi) = A\cos(\phi) + B\sin(\phi).
\end{equation}
Now compute $A$ and $B$ as
\begin{equation}
A = \int_{-\pi}^{+\pi}V_r(\phi)\cos(\phi)d\phi,\quad B = \int_{-\pi}^{+\pi}V_r(\phi)\sin(\phi)d\phi,
\end{equation}
then
\begin{equation}
V_{r0}(r) = \sqrt{A^2+B^2},\quad \phi_{r0}(r) = \tan^{-1}(B/A).
\end{equation}

\end{appendix}

\end{document}